\documentclass[11pt,a4paper]{article}
\pdfoutput=1
\usepackage{jcappub}
\usepackage{cprotect}
\usepackage{float}
\usepackage{subfiles}

\usepackage[symbol]{footmisc}

\newcommand{\PRE}[1]{{#1}} % Use if preprint style

\newcommand{\be}{\begin{equation}}
\newcommand{\ee}{\end{equation}}
\newcommand{\bea}{\begin{eqnarray}}
\newcommand{\eea}{\end{eqnarray}}

\def\gev{\text{ GeV}}
\def\mev{\text{ MeV}}
\def\kev{\text{ keV}}

\def\pb{\text{ pb}}

\def\kT{\text{ kT}}
\def\sr{\text{ sr}}
\def\cm{\text{ cm}}
\def\m{\text{ m}}

\def\s{\text{ s}}
\def\yr{\text{ yr}}

\newcommand{\Ar}[1]{{}^{#1}\!\text{Ar}}
\def\beq{\begin{eqnarray}}
\def\eeq{\end{eqnarray}}
\def\bea{\begin{eqnarray}}
\def\eea{\end{eqnarray}}

\newcommand{\gsim}{\lower.7ex\hbox{$\;\stackrel{\textstyle>}{\sim}\;$}}
\newcommand{\lsim}{\lower.7ex\hbox{$\;\stackrel{\textstyle<}{\sim}\;$}}

\usepackage{enumerate}
\include{epsf}

\begin{document}

\title{
\textsc{\textcolor{black}{Searching for Solar KDAR with DUNE}}
\PRE{\vspace*{0.1in}}
}

\newcommand{\Abilene}{Abilene Christian University, Abilene, TX 79601, USA}

\newcommand{\Albanysuny}{University of Albany, SUNY, Albany, NY 12222, USA}

\newcommand{\Amsterdam}{University of Amsterdam, NL-1098 XG Amsterdam, The Netherlands}

\newcommand{\Antalya}{Antalya Bilim University, 07190 D{\"o}{\c{s}}emealt{\i}/Antalya, Turkey}

\newcommand{\Antananarivo}{University of Antananarivo, Antananarivo 101, Madagascar}

\newcommand{\AntonioNarino}{Universidad Antonio Nari{\~n}o, Bogot{\'a}, Colombia}

\newcommand{\Argonne}{Argonne National Laboratory, Argonne, IL 60439, USA}

\newcommand{\Arizona}{University of Arizona, Tucson, AZ 85721, USA}

\newcommand{\Asuncion}{Universidad Nacional de Asunci{\'o}n, San Lorenzo, Paraguay}

\newcommand{\Athens}{University of Athens, Zografou GR 157 84, Greece}

\newcommand{\Atlantico}{Universidad del Atl{\'a}ntico, Barranquilla, Atl{\'a}ntico, Colombia}

\newcommand{\Augustana}{Augustana University, Sioux Falls, SD 57197, USA}

\newcommand{\Banaras}{Banaras Hindu University, Varanasi - 221 005, India}

\newcommand{\Basel}{University of Basel, CH-4056 Basel, Switzerland}

\newcommand{\Bern}{University of Bern, CH-3012 Bern, Switzerland}

\newcommand{\Beykent}{Beykent University, Istanbul, Turkey}

\newcommand{\Birmingham}{University of Birmingham, Birmingham B15 2TT, United Kingdom}

\newcommand{\BolognaUniversity}{Universit{\`a} del Bologna, 40127 Bologna, Italy}

\newcommand{\Boston}{Boston University, Boston, MA 02215, USA}

\newcommand{\Bristol}{University of Bristol, Bristol BS8 1TL, United Kingdom}

\newcommand{\Brookhaven}{Brookhaven National Laboratory, Upton, NY 11973, USA}

\newcommand{\Bucharest}{University of Bucharest, Bucharest, Romania}

\newcommand{\CBPF}{Centro Brasileiro de Pesquisas F\'isicas, Rio de Janeiro, RJ 22290-180, Brazil}

\newcommand{\CEASaclay}{IRFU, CEA, Universit{\'e} Paris-Saclay, F-91191 Gif-sur-Yvette, France}

\newcommand{\CERN}{CERN, The European Organization for Nuclear Research, 1211 Meyrin, Switzerland}

\newcommand{\CIEMAT}{CIEMAT, Centro de Investigaciones Energ{\'e}ticas, Medioambientales y Tecnol{\'o}gicas, E-28040 Madrid, Spain}

\newcommand{\CUSB}{Central University of South Bihar, Gaya, 824236, India }

\newcommand{\CalBerkeley}{University of California Berkeley, Berkeley, CA 94720, USA}

\newcommand{\CalDavis}{University of California Davis, Davis, CA 95616, USA}

\newcommand{\CalIrvine}{University of California Irvine, Irvine, CA 92697, USA}

\newcommand{\CalLosangeles}{University of California Los Angeles, Los Angeles, CA 90095, USA}

\newcommand{\CalRiverside}{University of California Riverside, Riverside CA 92521, USA}

\newcommand{\CalSantabarbara}{University of California Santa Barbara, Santa Barbara, California 93106 USA}

\newcommand{\Caltech}{California Institute of Technology, Pasadena, CA 91125, USA}

\newcommand{\Cambridge}{University of Cambridge, Cambridge CB3 0HE, United Kingdom}

\newcommand{\Campinas}{Universidade Estadual de Campinas, Campinas - SP, 13083-970, Brazil}

\newcommand{\CataniaUniversitadi}{Universit{\`a} di Catania, 2 - 95131 Catania, Italy}

\newcommand{\Charles}{Institute of Particle and Nuclear Physics of the Faculty of Mathematics and Physics of the Charles University, 180 00 Prague 8, Czech Republic }

\newcommand{\Chicago}{University of Chicago, Chicago, IL 60637, USA}

\newcommand{\ChungAng}{Chung-Ang University, Seoul 06974, South Korea}

\newcommand{\Cincinnati}{University of Cincinnati, Cincinnati, OH 45221, USA}

\newcommand{\Cinvestav}{Centro de Investigaci{\'o}n y de Estudios Avanzados del Instituto Polit{\'e}cnico Nacional (Cinvestav), Mexico City, Mexico}

\newcommand{\Colima}{Universidad de Colima, Colima, Mexico}

\newcommand{\ColoradoBoulder}{University of Colorado Boulder, Boulder, CO 80309, USA}

\newcommand{\ColoradoState}{Colorado State University, Fort Collins, CO 80523, USA}

\newcommand{\Columbia}{Columbia University, New York, NY 10027, USA}

\newcommand{\CzechAcademyofSciences}{Institute of Physics, Czech Academy of Sciences, 182 00 Prague 8, Czech Republic}

\newcommand{\CzechTechnical}{Czech Technical University, 115 19 Prague 1, Czech Republic}

\newcommand{\DakotaState}{Dakota State University, Madison, SD 57042, USA}

\newcommand{\Dallas}{University of Dallas, Irving, TX 75062-4736, USA}

\newcommand{\DannecyleVieux}{Laboratoire d{\textquoteright}Annecy de Physique des Particules, Univ. Grenoble Alpes, Univ. Savoie Mont Blanc, CNRS, LAPP-IN2P3, 74000 Annecy, France}

\newcommand{\Daresbury}{Daresbury Laboratory, Cheshire WA4 4AD, United Kingdom}

\newcommand{\Drexel}{Drexel University, Philadelphia, PA 19104, USA}

\newcommand{\Duke}{Duke University, Durham, NC 27708, USA}

\newcommand{\Durham}{Durham University, Durham DH1 3LE, United Kingdom}

\newcommand{\EIA}{Universidad EIA, Envigado, Antioquia, Colombia}

\newcommand{\ETH}{ETH Zurich, Zurich, Switzerland}

\newcommand{\Edinburgh}{University of Edinburgh, Edinburgh EH8 9YL, United Kingdom}

\newcommand{\FCULport}{Faculdade de Ci{\^e}ncias da Universidade de Lisboa - FCUL, 1749-016 Lisboa, Portugal}

\newcommand{\FederaldeAlfenas}{Universidade Federal de Alfenas, Po{\c{c}}os de Caldas - MG, 37715-400, Brazil}

\newcommand{\FederaldeGoias}{Universidade Federal de Goias, Goiania, GO 74690-900, Brazil}

\newcommand{\FederaldeSaoCarlos}{Universidade Federal de S{\~a}o Carlos, Araras - SP, 13604-900, Brazil}

\newcommand{\FederaldoABC}{Universidade Federal do ABC, Santo Andr{\'e} - SP, 09210-580, Brazil}

\newcommand{\FederaldoRio}{Universidade Federal do Rio de Janeiro,  Rio de Janeiro - RJ, 21941-901, Brazil}

\newcommand{\Fermi}{Fermi National Accelerator Laboratory, Batavia, IL 60510, USA}

\newcommand{\Ferrarauniv}{University of Ferrara, Ferrara, Italy}

\newcommand{\Florida}{University of Florida, Gainesville, FL 32611-8440, USA}

\newcommand{\Fluminense}{Fluminense Federal University, 9 Icara{\'\i} Niter{\'o}i - RJ, 24220-900, Brazil }

\newcommand{\Genova}{Universit{\`a} degli Studi di Genova, Genova, Italy}

\newcommand{\Georgian}{Georgian Technical University, Tbilisi, Georgia}

\newcommand{\GranSasso}{Gran Sasso Science Institute, L'Aquila, Italy}

\newcommand{\GranSassoLab}{Laboratori Nazionali del Gran Sasso, L'Aquila AQ, Italy}

\newcommand{\Granada}{University of Granada {\&} CAFPE, 18002 Granada, Spain}

\newcommand{\Grenoble}{University Grenoble Alpes, CNRS, Grenoble INP, LPSC-IN2P3, 38000 Grenoble, France}

\newcommand{\Guanajuato}{Universidad de Guanajuato, Guanajuato, C.P. 37000, Mexico}

\newcommand{\Harish}{Harish-Chandra Research Institute, Jhunsi, Allahabad 211 019, India}

\newcommand{\Harvard}{Harvard University, Cambridge, MA 02138, USA}

\newcommand{\Hawaii}{University of Hawaii, Honolulu, HI 96822, USA}

\newcommand{\Houston}{University of Houston, Houston, TX 77204, USA}

\newcommand{\Hyderabad}{University of  Hyderabad, Gachibowli, Hyderabad - 500 046, India}

\newcommand{\IFAE}{Institut de F{\'\i}sica d{\textquoteright}Altes Energies (IFAE){\textemdash}Barcelona Institute of Science and Technology (BIST), Barcelona, Spain}

\newcommand{\IFIC}{Instituto de F{\'\i}sica Corpuscular, CSIC and Universitat de Val{\`e}ncia, 46980 Paterna, Valencia, Spain}

\newcommand{\IGFAE}{Instituto Galego de Fisica de Altas Enerxias, A Coru{\~n}a, Spain}

\newcommand{\INFNBologna}{Istituto Nazionale di Fisica Nucleare Sezione di Bologna, 40127 Bologna BO, Italy}

\newcommand{\INFNCatania}{Istituto Nazionale di Fisica Nucleare Sezione di Catania, I-95123 Catania, Italy}

\newcommand{\INFNFerrara}{Istituto Nazionale di Fisica Nucleare Sezione di Ferrara, I-44122 Ferrara, Italy}

\newcommand{\INFNGenova}{Istituto Nazionale di Fisica Nucleare Sezione di Genova, 16146 Genova GE, Italy}

\newcommand{\INFNLecce}{Istituto Nazionale di Fisica Nucleare Sezione di Lecce, 73100 - Lecce, Italy}

\newcommand{\INFNMilanBicocca}{Istituto Nazionale di Fisica Nucleare Sezione di Milano Bicocca, 3 - I-20126 Milano, Italy}

\newcommand{\INFNMilano}{Istituto Nazionale di Fisica Nucleare Sezione di Milano, 20133 Milano, Italy}

\newcommand{\INFNNapoli}{Istituto Nazionale di Fisica Nucleare Sezione di Napoli, I-80126 Napoli, Italy}

\newcommand{\INFNPadova}{Istituto Nazionale di Fisica Nucleare Sezione di Padova, 35131 Padova, Italy}

\newcommand{\INFNPavia}{Istituto Nazionale di Fisica Nucleare Sezione di Pavia,  I-27100 Pavia, Italy}

\newcommand{\INFNSud}{Istituto Nazionale di Fisica Nucleare Laboratori Nazionali del Sud, 95123 Catania, Italy}

\newcommand{\INR}{Institute for Nuclear Research of the Russian Academy of Sciences, Moscow 117312, Russia}

\newcommand{\IPLyon}{Institut de Physique des 2 Infinis de Lyon, 69622 Villeurbanne, France}

\newcommand{\IPM}{Institute for Research in Fundamental Sciences, Tehran, Iran}

\newcommand{\ISTlisboa}{Instituto Superior T{\'e}cnico - IST, Universidade de Lisboa, Portugal}

\newcommand{\Idaho}{Idaho State University, Pocatello, ID 83209, USA}

\newcommand{\Illinoisinstitute}{Illinois Institute of Technology, Chicago, IL 60616, USA}

\newcommand{\Imperial}{Imperial College of Science Technology and Medicine, London SW7 2BZ, United Kingdom}

\newcommand{\IndGuwahati}{Indian Institute of Technology Guwahati, Guwahati, 781 039, India}

\newcommand{\IndHyderabad}{Indian Institute of Technology Hyderabad, Hyderabad, 502285, India}

\newcommand{\Indiana}{Indiana University, Bloomington, IN 47405, USA}

\newcommand{\Ingenieria}{Universidad Nacional de Ingenier{\'\i}a, Lima 25, Per{\'u}}

\newcommand{\Insubria }{University of Insubria, Via Ravasi, 2, 21100 Varese VA, Italy}

\newcommand{\Iowa}{University of Iowa, Iowa City, IA 52242, USA}

\newcommand{\IowaState}{Iowa State University, Ames, Iowa 50011, USA}

\newcommand{\Iwate}{Iwate University, Morioka, Iwate 020-8551, Japan}

\newcommand{\JINR}{Joint Institute for Nuclear Research, Dzhelepov Laboratory of Nuclear Problems 6 Joliot-Curie, Dubna, Moscow Region, 141980 RU }

\newcommand{\Jammu}{University of Jammu, Jammu-180006, India}

\newcommand{\Jawaharlal}{Jawaharlal Nehru University, New Delhi 110067, India}

\newcommand{\Jeonbuk}{Jeonbuk National University, Jeonrabuk-do 54896, South Korea}

\newcommand{\Jyvaskyla}{University of Jyvaskyla, FI-40014, Finland}

\newcommand{\KEK}{High Energy Accelerator Research Organization (KEK), Ibaraki, 305-0801, Japan}

\newcommand{\KISTI}{Korea Institute of Science and Technology Information, Daejeon, 34141, South Korea}

\newcommand{\KL}{K L University, Vaddeswaram, Andhra Pradesh 522502, India}

\newcommand{\Kansasstate}{Kansas State University, Manhattan, KS 66506, USA}

\newcommand{\Kavli}{Kavli Institute for the Physics and Mathematics of the Universe, Kashiwa, Chiba 277-8583, Japan}

\newcommand{\Kure}{National Institute of Technology, Kure College, Hiroshima, 737-8506, Japan}

\newcommand{\Kyiv}{Taras Shevchenko National University of Kyiv, 01601 Kyiv, Ukraine}

\newcommand{\LIP}{Laborat{\'o}rio de Instrumenta{\c{c}}{\~a}o e F{\'\i}sica Experimental de Part{\'\i}culas, 1649-003 Lisboa and 3004-516 Coimbra, Portugal}

\newcommand{\Lancaster}{Lancaster University, Lancaster LA1 4YB, United Kingdom}

\newcommand{\LawrenceBerkeley}{Lawrence Berkeley National Laboratory, Berkeley, CA 94720, USA}

\newcommand{\Liverpool}{University of Liverpool, L69 7ZE, Liverpool, United Kingdom}

\newcommand{\LosAlmos}{Los Alamos National Laboratory, Los Alamos, NM 87545, USA}

\newcommand{\Louisanastate}{Louisiana State University, Baton Rouge, LA 70803, USA}

\newcommand{\Lucknow}{University of Lucknow, Uttar Pradesh 226007, India}

\newcommand{\Madrid}{Madrid Autonoma University and IFT UAM/CSIC, 28049 Madrid, Spain}

\newcommand{\Manchester}{University of Manchester, Manchester M13 9PL, United Kingdom}

\newcommand{\Massinsttech}{Massachusetts Institute of Technology, Cambridge, MA 02139, USA}

\newcommand{\Maxplanck}{Max-Planck-Institut, Munich, 80805, Germany}

\newcommand{\Medellin}{University of Medell{\'\i}n, Medell{\'\i}n, 050026 Colombia }

\newcommand{\Michigan}{University of Michigan, Ann Arbor, MI 48109, USA}

\newcommand{\Michiganstate}{Michigan State University, East Lansing, MI 48824, USA}

\newcommand{\MilanoBicocca}{Universit{\`a} del Milano-Bicocca, 20126 Milano, Italy}

\newcommand{\MilanoUniv}{Universit{\`a} degli Studi di Milano, I-20133 Milano, Italy}

\newcommand{\Minnduluth}{University of Minnesota Duluth, Duluth, MN 55812, USA}

\newcommand{\Minntwin}{University of Minnesota Twin Cities, Minneapolis, MN 55455, USA}

\newcommand{\Mississippi}{University of Mississippi, University, MS 38677 USA}

\newcommand{\Newmexico}{University of New Mexico, Albuquerque, NM 87131, USA}

\newcommand{\Niewodniczanski}{H. Niewodnicza{\'n}ski Institute of Nuclear Physics, Polish Academy of Sciences, Cracow, Poland}

\newcommand{\Nikhef}{Nikhef National Institute of Subatomic Physics, 1098 XG Amsterdam, Netherlands}

\newcommand{\Northdakota}{University of North Dakota, Grand Forks, ND 58202-8357, USA}

\newcommand{\Northernillinois}{Northern Illinois University, DeKalb, IL 60115, USA}

\newcommand{\Northwestern}{Northwestern University, Evanston, Il 60208, USA}

\newcommand{\NotreDame}{University of Notre Dame, Notre Dame, IN 46556, USA}

\newcommand{\Occidental}{Occidental College, Los Angeles, CA  90041}

\newcommand{\Ohiostate}{Ohio State University, Columbus, OH 43210, USA}

\newcommand{\OregonState}{Oregon State University, Corvallis, OR 97331, USA}

\newcommand{\Oxford}{University of Oxford, Oxford, OX1 3RH, United Kingdom}

\newcommand{\PacificNorthwest}{Pacific Northwest National Laboratory, Richland, WA 99352, USA}

\newcommand{\Padova}{Universt{\`a} degli Studi di Padova, I-35131 Padova, Italy}

\newcommand{\Panjab}{Panjab University, Chandigarh, 160014 U.T., India}

\newcommand{\Parissaclay}{Universit{\'e} Paris-Saclay, CNRS/IN2P3, IJCLab, 91405 Orsay, France}

\newcommand{\Parisuniversite}{Universit{\'e} de Paris, CNRS, Astroparticule et Cosmologie, F-75006, Paris, France}

\newcommand{\Pavia}{Universit{\`a} degli Studi di Pavia, 27100 Pavia PV, Italy}

\newcommand{\Penn}{University of Pennsylvania, Philadelphia, PA 19104, USA}

\newcommand{\PennState}{Pennsylvania State University, University Park, PA 16802, USA}

\newcommand{\PhysicalResearchLaboratory}{Physical Research Laboratory, Ahmedabad 380 009, India}

\newcommand{\Pisa}{Universit{\`a} di Pisa, I-56127 Pisa, Italy}

\newcommand{\Pitt}{University of Pittsburgh, Pittsburgh, PA 15260, USA}

\newcommand{\Pontificia}{Pontificia Universidad Cat{\'o}lica del Per{\'u}, Lima, Per{\'u}}

\newcommand{\PuertoRico}{University of Puerto Rico, Mayaguez 00681, Puerto Rico, USA}

\newcommand{\Punjab}{Punjab Agricultural University, Ludhiana 141004, India}

\newcommand{\QMUL}{Queen Mary University of London, London E1 4NS, United Kingdom }

\newcommand{\Radboud}{Radboud University, NL-6525 AJ Nijmegen, Netherlands}

\newcommand{\Rochester}{University of Rochester, Rochester, NY 14627, USA}

\newcommand{\Royalholloway}{Royal Holloway College London, TW20 0EX, United Kingdom}

\newcommand{\Rutgers}{Rutgers University, Piscataway, NJ, 08854, USA}

\newcommand{\Rutherford}{STFC Rutherford Appleton Laboratory, Didcot OX11 0QX, United Kingdom}

\newcommand{\SLAC}{SLAC National Accelerator Laboratory, Menlo Park, CA 94025, USA}

\newcommand{\SURF}{Sanford Underground Research Facility, Lead, SD, 57754, USA}

\newcommand{\Salento}{Universit{\`a} del Salento, 73100 Lecce, Italy}

\newcommand{\Sanjosestate}{San Jose State University, San Jos{\'e}, CA 95192-0106, USA}

\newcommand{\SergioArboleda}{Universidad Sergio Arboleda, 11022 Bogot{\'a}, Colombia}

\newcommand{\Sheffield}{University of Sheffield, Sheffield S3 7RH, United Kingdom}

\newcommand{\SouthDakotaSchool}{South Dakota School of Mines and Technology, Rapid City, SD 57701, USA}

\newcommand{\SouthDakotaState}{South Dakota State University, Brookings, SD 57007, USA}

\newcommand{\Southcarolina}{University of South Carolina, Columbia, SC 29208, USA}

\newcommand{\SouthernMethodist}{Southern Methodist University, Dallas, TX 75275, USA}

\newcommand{\StonyBrook}{Stony Brook University, SUNY, Stony Brook, NY 11794, USA}

\newcommand{\Sungkyunkwan}{Sungkyunkwan University, Suwon, 16419, Korea, visitor to the collaboration}

\newcommand{\Sunyatsen}{Sun Yat-Sen University, Guangzhou, 510275}

\newcommand{\Sussex}{University of Sussex, Brighton, BN1 9RH, United Kingdom}

\newcommand{\Syracuse}{Syracuse University, Syracuse, NY 13244, USA}

\newcommand{\Tecnologica }{Universidade Tecnol{\'o}gica Federal do Paran{\'a}, Curitiba, Brazil}

\newcommand{\TexasAMcollege}{Texas A{\&}M University, College Station, Texas 77840}

\newcommand{\TexasAMcorpuscristi}{Texas A{\&}M University - Corpus Christi, Corpus Christi, TX 78412, USA}

\newcommand{\TexasArlington}{University of Texas at Arlington, Arlington, TX 76019, USA}

\newcommand{\Texasaustin}{University of Texas at Austin, Austin, TX 78712, USA}

\newcommand{\Toronto}{University of Toronto, Toronto, Ontario M5S 1A1, Canada}

\newcommand{\Tufts}{Tufts University, Medford, MA 02155, USA}

\newcommand{\UNIST}{Ulsan National Institute of Science and Technology, Ulsan 689-798, South Korea}

\newcommand{\Unifesp}{Universidade Federal de S{\~a}o Paulo, 09913-030, S{\~a}o Paulo, Brazil}

\newcommand{\UniversityCollegeLondon}{University College London, London, WC1E 6BT, United Kingdom}

%
% For exceptional author C. Rott
%
\newcommand{\Utah}{University of Utah, Salt Lake City, UT 84112, USA, visitor to the collaboration}

\newcommand{\ValleyCity}{Valley City State University, Valley City, ND 58072, USA}

\newcommand{\VariableEnergy}{Variable Energy Cyclotron Centre, 700 064 West Bengal, India}

\newcommand{\VirginiaTech}{Virginia Tech, Blacksburg, VA 24060, USA}

\newcommand{\Warsaw}{University of Warsaw, 02-093 Warsaw, Poland}

\newcommand{\Warwick}{University of Warwick, Coventry CV4 7AL, United Kingdom}

\newcommand{\Wellesley}{Wellesley College, Wellesley, MA 02481, USA}

\newcommand{\Wichita}{Wichita State University, Wichita, KS 67260, USA}

\newcommand{\WilliamMary}{William and Mary, Williamsburg, VA 23187, USA}

\newcommand{\Wisconsin}{University of Wisconsin Madison, Madison, WI 53706, USA}

\newcommand{\Yale}{Yale University, New Haven, CT 06520, USA}

\newcommand{\Yerevan}{Yerevan Institute for Theoretical Physics and Modeling, Yerevan 0036, Armenia}

\newcommand{\York}{York University, Toronto M3J 1P3, Canada}

%----------------------------------------------------

% So that institutions appear in alphabetical order:

\newcommand{\iAbilene}{1} \affiliation[\iAbilene]{\Abilene}
\newcommand{\iAlbanysuny}{2} \affiliation[\iAlbanysuny]{\Albanysuny}
\newcommand{\iAmsterdam}{3} \affiliation[\iAmsterdam]{\Amsterdam}
\newcommand{\iAntalya}{4} \affiliation[\iAntalya]{\Antalya}
\newcommand{\iAntananarivo}{5} \affiliation[\iAntananarivo]{\Antananarivo}
\newcommand{\iAntonioNarino}{6} \affiliation[\iAntonioNarino]{\AntonioNarino}
\newcommand{\iArgonne}{7} \affiliation[\iArgonne]{\Argonne}
\newcommand{\iArizona}{8} \affiliation[\iArizona]{\Arizona}
\newcommand{\iAsuncion}{9} \affiliation[\iAsuncion]{\Asuncion}
\newcommand{\iAthens}{10} \affiliation[\iAthens]{\Athens}
\newcommand{\iAtlantico}{11} \affiliation[\iAtlantico]{\Atlantico}
\newcommand{\iAugustana}{12} \affiliation[\iAugustana]{\Augustana}
\newcommand{\iBanaras}{13} \affiliation[\iBanaras]{\Banaras}
\newcommand{\iBasel}{14} \affiliation[\iBasel]{\Basel}
\newcommand{\iBern}{15} \affiliation[\iBern]{\Bern}
\newcommand{\iBeykent}{16} \affiliation[\iBeykent]{\Beykent}
\newcommand{\iBirmingham}{17} \affiliation[\iBirmingham]{\Birmingham}
\newcommand{\iBolognaUniversity}{18} \affiliation[\iBolognaUniversity]{\BolognaUniversity}
\newcommand{\iBoston}{19} \affiliation[\iBoston]{\Boston}
\newcommand{\iBristol}{20} \affiliation[\iBristol]{\Bristol}
\newcommand{\iBrookhaven}{21} \affiliation[\iBrookhaven]{\Brookhaven}
\newcommand{\iBucharest}{22} \affiliation[\iBucharest]{\Bucharest}
\newcommand{\iCBPF}{23} \affiliation[\iCBPF]{\CBPF}
\newcommand{\iCEASaclay}{24} \affiliation[\iCEASaclay]{\CEASaclay}
\newcommand{\iCERN}{25} \affiliation[\iCERN]{\CERN}
\newcommand{\iCIEMAT}{26} \affiliation[\iCIEMAT]{\CIEMAT}
\newcommand{\iCUSB}{27} \affiliation[\iCUSB]{\CUSB}
\newcommand{\iCalBerkeley}{28} \affiliation[\iCalBerkeley]{\CalBerkeley}
\newcommand{\iCalDavis}{29} \affiliation[\iCalDavis]{\CalDavis}
\newcommand{\iCalIrvine}{30} \affiliation[\iCalIrvine]{\CalIrvine}
\newcommand{\iCalLosangeles}{31} \affiliation[\iCalLosangeles]{\CalLosangeles}
\newcommand{\iCalRiverside}{32} \affiliation[\iCalRiverside]{\CalRiverside}
\newcommand{\iCalSantabarbara}{33} \affiliation[\iCalSantabarbara]{\CalSantabarbara}
\newcommand{\iCaltech}{34} \affiliation[\iCaltech]{\Caltech}
\newcommand{\iCambridge}{35} \affiliation[\iCambridge]{\Cambridge}
\newcommand{\iCampinas}{36} \affiliation[\iCampinas]{\Campinas}
\newcommand{\iCataniaUniversitadi}{37} \affiliation[\iCataniaUniversitadi]{\CataniaUniversitadi}
\newcommand{\iCharles}{38} \affiliation[\iCharles]{\Charles}
\newcommand{\iChicago}{39} \affiliation[\iChicago]{\Chicago}
\newcommand{\iChungAng}{40} \affiliation[\iChungAng]{\ChungAng}
\newcommand{\iCincinnati}{41} \affiliation[\iCincinnati]{\Cincinnati}
\newcommand{\iCinvestav}{42} \affiliation[\iCinvestav]{\Cinvestav}
\newcommand{\iColima}{43} \affiliation[\iColima]{\Colima}
\newcommand{\iColoradoBoulder}{44} \affiliation[\iColoradoBoulder]{\ColoradoBoulder}
\newcommand{\iColoradoState}{45} \affiliation[\iColoradoState]{\ColoradoState}
\newcommand{\iColumbia}{46} \affiliation[\iColumbia]{\Columbia}
\newcommand{\iCzechAcademyofSciences}{47} \affiliation[\iCzechAcademyofSciences]{\CzechAcademyofSciences}
\newcommand{\iCzechTechnical}{48} \affiliation[\iCzechTechnical]{\CzechTechnical}
\newcommand{\iDakotaState}{49} \affiliation[\iDakotaState]{\DakotaState}
\newcommand{\iDallas}{50} \affiliation[\iDallas]{\Dallas}
\newcommand{\iDannecyleVieux}{51} \affiliation[\iDannecyleVieux]{\DannecyleVieux}
\newcommand{\iDaresbury}{52} \affiliation[\iDaresbury]{\Daresbury}
\newcommand{\iDrexel}{53} \affiliation[\iDrexel]{\Drexel}
\newcommand{\iDuke}{54} \affiliation[\iDuke]{\Duke}
\newcommand{\iDurham}{55} \affiliation[\iDurham]{\Durham}
\newcommand{\iEIA}{56} \affiliation[\iEIA]{\EIA}
\newcommand{\iETH}{57} \affiliation[\iETH]{\ETH}
\newcommand{\iEdinburgh}{58} \affiliation[\iEdinburgh]{\Edinburgh}
\newcommand{\iFCULport}{59} \affiliation[\iFCULport]{\FCULport}
\newcommand{\iFederaldeAlfenas}{60} \affiliation[\iFederaldeAlfenas]{\FederaldeAlfenas}
\newcommand{\iFederaldeGoias}{61} \affiliation[\iFederaldeGoias]{\FederaldeGoias}
\newcommand{\iFederaldeSaoCarlos}{62} \affiliation[\iFederaldeSaoCarlos]{\FederaldeSaoCarlos}
\newcommand{\iFederaldoABC}{63} \affiliation[\iFederaldoABC]{\FederaldoABC}
\newcommand{\iFederaldoRio}{64} \affiliation[\iFederaldoRio]{\FederaldoRio}
\newcommand{\iFermi}{65} \affiliation[\iFermi]{\Fermi}
\newcommand{\iFerrarauniv}{66} \affiliation[\iFerrarauniv]{\Ferrarauniv}
\newcommand{\iFlorida}{67} \affiliation[\iFlorida]{\Florida}
\newcommand{\iFluminense}{68} \affiliation[\iFluminense]{\Fluminense}
\newcommand{\iGenova}{69} \affiliation[\iGenova]{\Genova}
\newcommand{\iGeorgian}{70} \affiliation[\iGeorgian]{\Georgian}
\newcommand{\iGranSasso}{71} \affiliation[\iGranSasso]{\GranSasso}
\newcommand{\iGranSassoLab}{72} \affiliation[\iGranSassoLab]{\GranSassoLab}
\newcommand{\iGranada}{73} \affiliation[\iGranada]{\Granada}
\newcommand{\iGrenoble}{74} \affiliation[\iGrenoble]{\Grenoble}
\newcommand{\iGuanajuato}{75} \affiliation[\iGuanajuato]{\Guanajuato}
\newcommand{\iHarish}{76} \affiliation[\iHarish]{\Harish}
\newcommand{\iHarvard}{77} \affiliation[\iHarvard]{\Harvard}
\newcommand{\iHawaii}{78} \affiliation[\iHawaii]{\Hawaii}
\newcommand{\iHouston}{79} \affiliation[\iHouston]{\Houston}
\newcommand{\iHyderabad}{80} \affiliation[\iHyderabad]{\Hyderabad}
\newcommand{\iIFAE}{81} \affiliation[\iIFAE]{\IFAE}
\newcommand{\iIFIC}{82} \affiliation[\iIFIC]{\IFIC}
\newcommand{\iIGFAE}{83} \affiliation[\iIGFAE]{\IGFAE}
\newcommand{\iINFNBologna}{84} \affiliation[\iINFNBologna]{\INFNBologna}
\newcommand{\iINFNCatania}{85} \affiliation[\iINFNCatania]{\INFNCatania}
\newcommand{\iINFNFerrara}{86} \affiliation[\iINFNFerrara]{\INFNFerrara}
\newcommand{\iINFNGenova}{87} \affiliation[\iINFNGenova]{\INFNGenova}
\newcommand{\iINFNLecce}{88} \affiliation[\iINFNLecce]{\INFNLecce}
\newcommand{\iINFNMilanBicocca}{89} \affiliation[\iINFNMilanBicocca]{\INFNMilanBicocca}
\newcommand{\iINFNMilano}{90} \affiliation[\iINFNMilano]{\INFNMilano}
\newcommand{\iINFNNapoli}{91} \affiliation[\iINFNNapoli]{\INFNNapoli}
\newcommand{\iINFNPadova}{92} \affiliation[\iINFNPadova]{\INFNPadova}
\newcommand{\iINFNPavia}{93} \affiliation[\iINFNPavia]{\INFNPavia}
\newcommand{\iINFNSud}{94} \affiliation[\iINFNSud]{\INFNSud}
\newcommand{\iINR}{95} \affiliation[\iINR]{\INR}
\newcommand{\iIPLyon}{96} \affiliation[\iIPLyon]{\IPLyon}
\newcommand{\iIPM}{97} \affiliation[\iIPM]{\IPM}
\newcommand{\iISTlisboa}{98} \affiliation[\iISTlisboa]{\ISTlisboa}
\newcommand{\iIdaho}{99} \affiliation[\iIdaho]{\Idaho}
\newcommand{\iIllinoisinstitute}{100} \affiliation[\iIllinoisinstitute]{\Illinoisinstitute}
\newcommand{\iImperial}{101} \affiliation[\iImperial]{\Imperial}
\newcommand{\iIndGuwahati}{102} \affiliation[\iIndGuwahati]{\IndGuwahati}
\newcommand{\iIndHyderabad}{103} \affiliation[\iIndHyderabad]{\IndHyderabad}
\newcommand{\iIndiana}{104} \affiliation[\iIndiana]{\Indiana}
\newcommand{\iIngenieria}{105} \affiliation[\iIngenieria]{\Ingenieria}
\newcommand{\iInsubria }{106} \affiliation[\iInsubria ]{\Insubria }
\newcommand{\iIowa}{107} \affiliation[\iIowa]{\Iowa}
\newcommand{\iIowaState}{108} \affiliation[\iIowaState]{\IowaState}
\newcommand{\iIwate}{109} \affiliation[\iIwate]{\Iwate}
\newcommand{\iJINR}{110} \affiliation[\iJINR]{\JINR}
\newcommand{\iJammu}{111} \affiliation[\iJammu]{\Jammu}
\newcommand{\iJawaharlal}{112} \affiliation[\iJawaharlal]{\Jawaharlal}
\newcommand{\iJeonbuk}{113} \affiliation[\iJeonbuk]{\Jeonbuk}
\newcommand{\iJyvaskyla}{114} \affiliation[\iJyvaskyla]{\Jyvaskyla}
\newcommand{\iKEK}{115} \affiliation[\iKEK]{\KEK}
\newcommand{\iKISTI}{116} \affiliation[\iKISTI]{\KISTI}
\newcommand{\iKL}{117} \affiliation[\iKL]{\KL}
\newcommand{\iKansasstate}{118} \affiliation[\iKansasstate]{\Kansasstate}
\newcommand{\iKavli}{119} \affiliation[\iKavli]{\Kavli}
\newcommand{\iKure}{120} \affiliation[\iKure]{\Kure}
\newcommand{\iKyiv}{121} \affiliation[\iKyiv]{\Kyiv}
\newcommand{\iLIP}{122} \affiliation[\iLIP]{\LIP}
\newcommand{\iLancaster}{123} \affiliation[\iLancaster]{\Lancaster}
\newcommand{\iLawrenceBerkeley}{124} \affiliation[\iLawrenceBerkeley]{\LawrenceBerkeley}
\newcommand{\iLiverpool}{125} \affiliation[\iLiverpool]{\Liverpool}
\newcommand{\iLosAlmos}{126} \affiliation[\iLosAlmos]{\LosAlmos}
\newcommand{\iLouisanastate}{127} \affiliation[\iLouisanastate]{\Louisanastate}
\newcommand{\iLucknow}{128} \affiliation[\iLucknow]{\Lucknow}
\newcommand{\iMadrid}{129} \affiliation[\iMadrid]{\Madrid}
\newcommand{\iManchester}{130} \affiliation[\iManchester]{\Manchester}
\newcommand{\iMassinsttech}{131} \affiliation[\iMassinsttech]{\Massinsttech}
\newcommand{\iMaxplanck}{132} \affiliation[\iMaxplanck]{\Maxplanck}
\newcommand{\iMedellin}{133} \affiliation[\iMedellin]{\Medellin}
\newcommand{\iMichigan}{134} \affiliation[\iMichigan]{\Michigan}
\newcommand{\iMichiganstate}{135} \affiliation[\iMichiganstate]{\Michiganstate}
\newcommand{\iMilanoBicocca}{136} \affiliation[\iMilanoBicocca]{\MilanoBicocca}
\newcommand{\iMilanoUniv}{137} \affiliation[\iMilanoUniv]{\MilanoUniv}
\newcommand{\iMinnduluth}{138} \affiliation[\iMinnduluth]{\Minnduluth}
\newcommand{\iMinntwin}{139} \affiliation[\iMinntwin]{\Minntwin}
\newcommand{\iMississippi}{140} \affiliation[\iMississippi]{\Mississippi}
\newcommand{\iNewmexico}{141} \affiliation[\iNewmexico]{\Newmexico}
\newcommand{\iNiewodniczanski}{142} \affiliation[\iNiewodniczanski]{\Niewodniczanski}
\newcommand{\iNikhef}{143} \affiliation[\iNikhef]{\Nikhef}
\newcommand{\iNorthdakota}{144} \affiliation[\iNorthdakota]{\Northdakota}
\newcommand{\iNorthernillinois}{145} \affiliation[\iNorthernillinois]{\Northernillinois}
\newcommand{\iNorthwestern}{146} \affiliation[\iNorthwestern]{\Northwestern}
\newcommand{\iNotreDame}{147} \affiliation[\iNotreDame]{\NotreDame}
\newcommand{\iOccidental}{148} \affiliation[\iOccidental]{\Occidental}
\newcommand{\iOhiostate}{149} \affiliation[\iOhiostate]{\Ohiostate}
\newcommand{\iOregonState}{150} \affiliation[\iOregonState]{\OregonState}
\newcommand{\iOxford}{151} \affiliation[\iOxford]{\Oxford}
\newcommand{\iPacificNorthwest}{152} \affiliation[\iPacificNorthwest]{\PacificNorthwest}
\newcommand{\iPadova}{153} \affiliation[\iPadova]{\Padova}
\newcommand{\iPanjab}{154} \affiliation[\iPanjab]{\Panjab}
\newcommand{\iParissaclay}{155} \affiliation[\iParissaclay]{\Parissaclay}
\newcommand{\iParisuniversite}{156} \affiliation[\iParisuniversite]{\Parisuniversite}
\newcommand{\iPavia}{157} \affiliation[\iPavia]{\Pavia}
\newcommand{\iPenn}{158} \affiliation[\iPenn]{\Penn}
\newcommand{\iPennState}{159} \affiliation[\iPennState]{\PennState}
\newcommand{\iPhysicalResearchLaboratory}{160} \affiliation[\iPhysicalResearchLaboratory]{\PhysicalResearchLaboratory}
\newcommand{\iPisa}{161} \affiliation[\iPisa]{\Pisa}
\newcommand{\iPitt}{162} \affiliation[\iPitt]{\Pitt}
\newcommand{\iPontificia}{163} \affiliation[\iPontificia]{\Pontificia}
\newcommand{\iPuertoRico}{164} \affiliation[\iPuertoRico]{\PuertoRico}
\newcommand{\iPunjab}{165} \affiliation[\iPunjab]{\Punjab}
\newcommand{\iQMUL}{166} \affiliation[\iQMUL]{\QMUL}
\newcommand{\iRadboud}{167} \affiliation[\iRadboud]{\Radboud}
\newcommand{\iRochester}{168} \affiliation[\iRochester]{\Rochester}
\newcommand{\iRoyalholloway}{169} \affiliation[\iRoyalholloway]{\Royalholloway}
\newcommand{\iRutgers}{170} \affiliation[\iRutgers]{\Rutgers}
\newcommand{\iRutherford}{171} \affiliation[\iRutherford]{\Rutherford}
\newcommand{\iSLAC}{172} \affiliation[\iSLAC]{\SLAC}
\newcommand{\iSURF}{173} \affiliation[\iSURF]{\SURF}
\newcommand{\iSalento}{174} \affiliation[\iSalento]{\Salento}
\newcommand{\iSanjosestate}{175} \affiliation[\iSanjosestate]{\Sanjosestate}
\newcommand{\iSergioArboleda}{176} \affiliation[\iSergioArboleda]{\SergioArboleda}
\newcommand{\iSheffield}{177} \affiliation[\iSheffield]{\Sheffield}
\newcommand{\iSouthDakotaSchool}{178} \affiliation[\iSouthDakotaSchool]{\SouthDakotaSchool}
\newcommand{\iSouthDakotaState}{179} \affiliation[\iSouthDakotaState]{\SouthDakotaState}
\newcommand{\iSouthcarolina}{180} \affiliation[\iSouthcarolina]{\Southcarolina}
\newcommand{\iSouthernMethodist}{181} \affiliation[\iSouthernMethodist]{\SouthernMethodist}
\newcommand{\iStonyBrook}{182} \affiliation[\iStonyBrook]{\StonyBrook}
\newcommand{\iSungkyunkwan}{183} \affiliation[\iSungkyunkwan]{\Sungkyunkwan}
\newcommand{\iSunyatsen}{184} \affiliation[\iSunyatsen]{\Sunyatsen}
\newcommand{\iSussex}{185} \affiliation[\iSussex]{\Sussex}
\newcommand{\iSyracuse}{186} \affiliation[\iSyracuse]{\Syracuse}
\newcommand{\iTecnologica }{187} \affiliation[\iTecnologica ]{\Tecnologica }
\newcommand{\iTexasAMcollege}{188} \affiliation[\iTexasAMcollege]{\TexasAMcollege}
\newcommand{\iTexasAMcorpuscristi}{189} \affiliation[\iTexasAMcorpuscristi]{\TexasAMcorpuscristi}
\newcommand{\iTexasArlington}{190} \affiliation[\iTexasArlington]{\TexasArlington}
\newcommand{\iTexasaustin}{191} \affiliation[\iTexasaustin]{\Texasaustin}
\newcommand{\iToronto}{192} \affiliation[\iToronto]{\Toronto}
\newcommand{\iTufts}{193} \affiliation[\iTufts]{\Tufts}
\newcommand{\iUNIST}{194} \affiliation[\iUNIST]{\UNIST}
\newcommand{\iUnifesp}{195} \affiliation[\iUnifesp]{\Unifesp}
\newcommand{\iUniversityCollegeLondon}{196} \affiliation[\iUniversityCollegeLondon]{\UniversityCollegeLondon}
%
% For exceptional author C. Rott. If accepted will be moved into
% alphabetical order.
%
\newcommand{\iUtah}{197} \affiliation[\iUtah]{\Utah}
\newcommand{\iValleyCity}{198} \affiliation[\iValleyCity]{\ValleyCity}
\newcommand{\iVariableEnergy}{199} \affiliation[\iVariableEnergy]{\VariableEnergy}
\newcommand{\iVirginiaTech}{200} \affiliation[\iVirginiaTech]{\VirginiaTech}
\newcommand{\iWarsaw}{201} \affiliation[\iWarsaw]{\Warsaw}
\newcommand{\iWarwick}{202} \affiliation[\iWarwick]{\Warwick}
\newcommand{\iWellesley}{203} \affiliation[\iWellesley]{\Wellesley}
\newcommand{\iWichita}{204} \affiliation[\iWichita]{\Wichita}
\newcommand{\iWilliamMary}{205} \affiliation[\iWilliamMary]{\WilliamMary}
\newcommand{\iWisconsin}{206} \affiliation[\iWisconsin]{\Wisconsin}
\newcommand{\iYale}{207} \affiliation[\iYale]{\Yale}
\newcommand{\iYerevan}{208} \affiliation[\iYerevan]{\Yerevan}
\newcommand{\iYork}{209} \affiliation[\iYork]{\York}

%----------------------------------------------------

% Authors in alphabetical order

\author[\iLiverpool,\iCERN]{A.~Abed Abud}
\author[\iOxford]{B.~Abi}
\author[\iFermi]{R.~Acciarri}
\author[\iAtlantico]{M.~A.~Acero}
\author[\iTecnologica ]{M.~R.~Adames}
\author[\iGeorgian]{G.~Adamov}
\author[\iBrookhaven]{D.~Adams}
\author[\iBristol]{M.~Adinolfi}
\author[\iHouston]{A.~Aduszkiewicz}
\author[\iLawrenceBerkeley]{J.~Aguilar}
% ...
%1179 authors

\collaboration{The DUNE Collaboration}

\emailAdd{\textcolor{black}{dune-www-pubs@fnal.gov}}

\abstract{\textcolor{black}{
The observation of 236 MeV muon neutrinos from kaon-decay-at-rest (KDAR) originating in the core of the Sun would provide a unique signature of dark matter annihilation. Since excellent angle and energy reconstruction are necessary to detect this monoenergetic, directional neutrino flux, DUNE with its vast volume and reconstruction capabilities, is a promising candidate for a KDAR neutrino search. In this work, we evaluate the proposed KDAR neutrino search strategies by realistically 
modeling both neutrino-nucleus interactions and the 
response of DUNE.  We find that, although reconstruction 
of the neutrino energy and direction is difficult with 
current techniques in the relevant energy range, the superb energy 
resolution, angular resolution, and particle identification 
offered by DUNE can still permit great signal/background 
discrimination.  Moreover, there are non-standard 
scenarios in which searches 
at DUNE
for KDAR in the Sun 
can probe dark matter interactions.}
}

\keywords{Dark matter, Solar WIMPs, indirect WIMP search}

\maketitle
\flushbottom
\date{\today}

%%%%%%%%%%%%%%%%%%%%%%%%%%%%%%%%%%%%%%%%%%%%%%%%%%%%%%%%%%

\section{Introduction}

There has been recent interest from the experimental community 
in detecting the neutrinos produced by kaon decay at rest 
(KDAR)~\cite{Spitz:2014hwa,Aguilar-Arevalo:2018ylq}.  
\textcolor{black}{One application }
for these techniques is the 
search for neutrinos produced when gravitationally-captured 
dark matter annihilates in the core of the Sun~\cite{Silk:1985ax,Press:1985ug,Krauss:1985ks}.  
\textcolor{black}{If dark matter annihilation produces u, d, and s quarks, then the result 
of subsequent hadronization and fragmentation 
would be a large number of $K^+$ which come to rest in the 
dense solar medium before decaying.  $64\%$ of these decays 
($K^+ \rightarrow \mu^+ \nu_\mu$) produce monoenergetic 
$\nu_\mu$ with an energy 
of $\sim 236 \mev$~\cite{Rott:2012qb,Bernal:2012qh,Rott:2015nma}.  The oscillations of these neutrinos while passing through the dense solar medium and vacuum results in 
approximately comparable fluxes of active neutrinos 
in all three 
flavors at Earth~\cite{Lehnert:2007fv}.} 
Recent work has 
focused on developing new techniques for utilizing 
the excellent 
particle identification and energy and angular resolution of 
DUNE to identify the energy and direction of the incoming 
$236 \mev$ neutrino~\cite{Rott:2016mzs}.  
The identification of a flux of 
$236 \mev$ neutrinos arriving from the Sun would be  
an extraordinary signal of new physics, providing a new 
handle on dark matter interactions which could be a unique 
probe of non-standard dark matter 
models~\cite{Rott:2017weo}.  This work further 
develops techniques for measuring the monoenergetic 
neutrinos arising from KDAR in the Sun, with a focus on 
increasing the signal-to-background ratio.

At water Cherenkov (WC) neutrino detectors, it 
is very difficult 
to determine the direction of an ${\cal O}(100) \mev$ 
neutrino because the charged lepton produced by a 
charged-current interaction is largely isotropic at these energies.  But 
in a large fraction of neutrino-argon CC-interactions, a 
proton is ejected preferentially in the forward direction.  
Though this proton cannot be seen in a WC detector, its 
energy and direction can be well-measured in a liquid argon 
time projection chamber (LArTPC) detector, 
\textcolor{black}{such as DUNE.  
Thus, although WC detectors will generally 
have a statistical advantage due to their size, LArTPC 
detectors can have an advantage in reducing some systematic 
uncertainties, due to a greater ability to reject 
background.}

In~\cite{Rott:2016mzs}, it was 
proposed that one search for DUNE events with exactly one proton and one 
charged lepton with a total energy of $236 \mev$, and with the proton directed away from 
the Sun.  It was found that this directionality strategy should improve DUNE sensitivity
to dark matter annihilation in the Sun, while yielding a signal-to-background 
ratio as high as $\sim 40\%$.  
\textcolor{black} {In this paper, we use 
%perform  a more realistic analysis, 
the LArSoft package~\cite{LArSoft} to realistically model 
the detector response, 
including the asymmetric response due to the orientation of the detector with respect 
to the incoming neutrino, and we use the Pandora package~\cite{Pandora} to perform track 
reconstruction.}
We also find that, although the charged lepton is produced roughly isotropically, its 
direction is correlated with that of the proton, providing a new method for rejecting 
background \textcolor{black}{that can significantly improve 
the signal-to-background ratio}.

At DUNE, the charged current interaction $\nu_\ell + \Ar{40} \rightarrow \ell^- + p^+ + \Ar{39}$ 
produces an ejected proton and charged lepton which can be 
well-measured~\cite{Acciarri:2015uup}.  But the recoil of the 
remnant $\Ar{39}$ will not be well-measured, and although the kinetic energy of the remnant 
nucleus will be small, its momentum may be substantial.  But given a hypothesis for the 
energy and momentum of the neutrino (i.~e., a $236 \mev$ neutrino arriving from the Sun), 
the momentum of the remnant nucleus can be reconstructed using momentum conservation. We find that when the struck proton is very forward-directed, 
the remnant nucleus is typically backscattered \textcolor{black}{(more on this in Section 2 and Fig.~\ref{fig:nucleon}).}   
Utilizing this correlation, we find that for models where evidence can be found 
at $90\%$ C.L. with a 400 kT yr exposure of DUNE, the signal-to-background ratio can be 
as high as $2.2$.

\textcolor{black}{We find that, with a 400 kT yr 
exposure, 
DUNE can probe ${\cal O}(10^3) \m^{-2}\s^{-1}$ fluxes of 
236 MeV $\nu_\mu$ emanating from the Sun.  As a specific 
example, we consider the case of low-mass dark matter 
($m \lesssim 10~\gev$) which scatters inelastically with nuclei.  
We estimate the sensitivity of DUNE to models which cannot be probed by direct 
detection experiments.
}

The plan of this paper is as follows.  In Section 2, we describe our simulation framework and analysis cuts.  In Section 3, we describe the resulting 
sensitivity to a flux of KDAR neutrinos, and as 
an example, interpret this as a sensitivity to 
a particular class of dark matter models which 
cannot be probed by direct detection 
experiments.  
We conclude with a discussion of 
our results in Section 4.

\section{Event Simulation and Analysis Cuts}

\textcolor{black}{Dark matter annihilation at the core 
of the Sun can produce light mesons, whose 
decays-at-rest can produce monoenergetic neutrinos.
KDAR ($K^+ \rightarrow \mu^+ \nu_\mu$) 
will produce a $E_\nu = 236 \mev$ monoenergetic $\nu_\mu$ at the
core of the Sun.  On the other hand, 
$K^-$ and $\pi^-$ will 
tend to be Coulomb-captured by nuclei. Hence the flux of neutrinos from $K^-$ and $\pi^-$ is 
small~\cite{Ponomarev:1973ya}. $\pi^+$ decay-at-rest in the Sun will produce a 
monoenergetic $30 \mev$ neutrino.  But this signal is less promising~\cite{Rott:2015nma}, 
because the background 
from atmospheric neutrinos is larger at these energies, while the $\nu-\Ar{40}$ cross section 
is smaller.  Moreover, the scattering of a $30 \mev$ neutrino is less likely to eject a proton, 
which is needed for directionality.  Dark matter annihilation can also produce muons 
which decay at rest, but this signal is less promising because it does not yield a 
monoenergetic neutrino.  As a result, we focus on the 
$236 \mev$ $\nu_\mu$ produced by KDAR in the Sun.}

\begin{figure}[hbt!]
  \centering
  \includegraphics[width= 1.0 \textwidth]{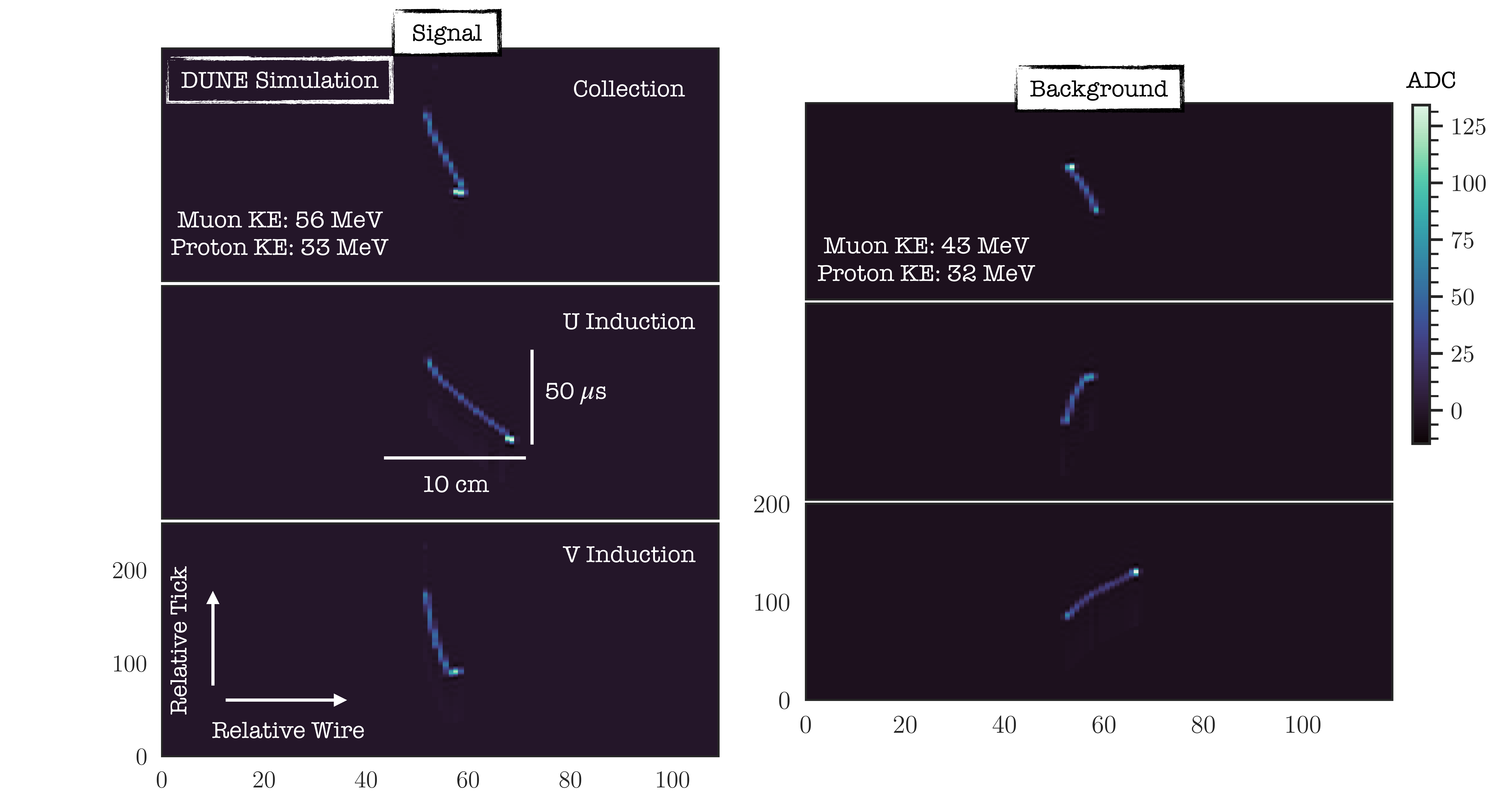}
  \cprotect\caption{\label{fig:evd_nu_mu} Time (ticks) vs. wire (number) view of a $236 \mev$ $\nu_\mu$ event simulated at DUNE. The color corresponds to the voltage read out on the wires (in ADCs). Each tick is 500 ns. Each panel corresponds to an individual wire plane. The wire spacing for the top (collection) plane is 4.79 mm. The wire spacing for the middle and bottom (induction) planes is 4.67 mm. The collection plane is aligned with the vertical of the detector frame and the induction planes are angled $35.7^{\circ}$ 
  with respect to vertical. A muon and a proton are ejected. \textcolor{black}{The muon is the longest track. Fig.~\ref{fig:KE} shows the distributions of the kinetic energies of the ejected protons and muons. The right panel shows a background event stemming from a neutrino of $190 \mev$.} }
\end{figure}

By the time this neutrino reaches Earth, it 
will have oscillated into all three flavors.  But 
only $\nu_\mu$ and $\nu_e$ can produce a 
charged-current interaction at this energy. In this analysis, we only consider $\nu_\mu$. 
\textcolor{black}{We are interested in charge-current 
events in which a muon is produced and a 
proton is ejected from the nucleus,}
\textcolor{black}{since these particles can leave 
crisp tracks in DUNE, as shown in Fig.~\ref{fig:evd_nu_mu}.}

%=============================================================================
%=============================================================================

%=============================================================================
\begin{figure}[hbt!]
  \centering
  \includegraphics[width= 0.8 \textwidth]{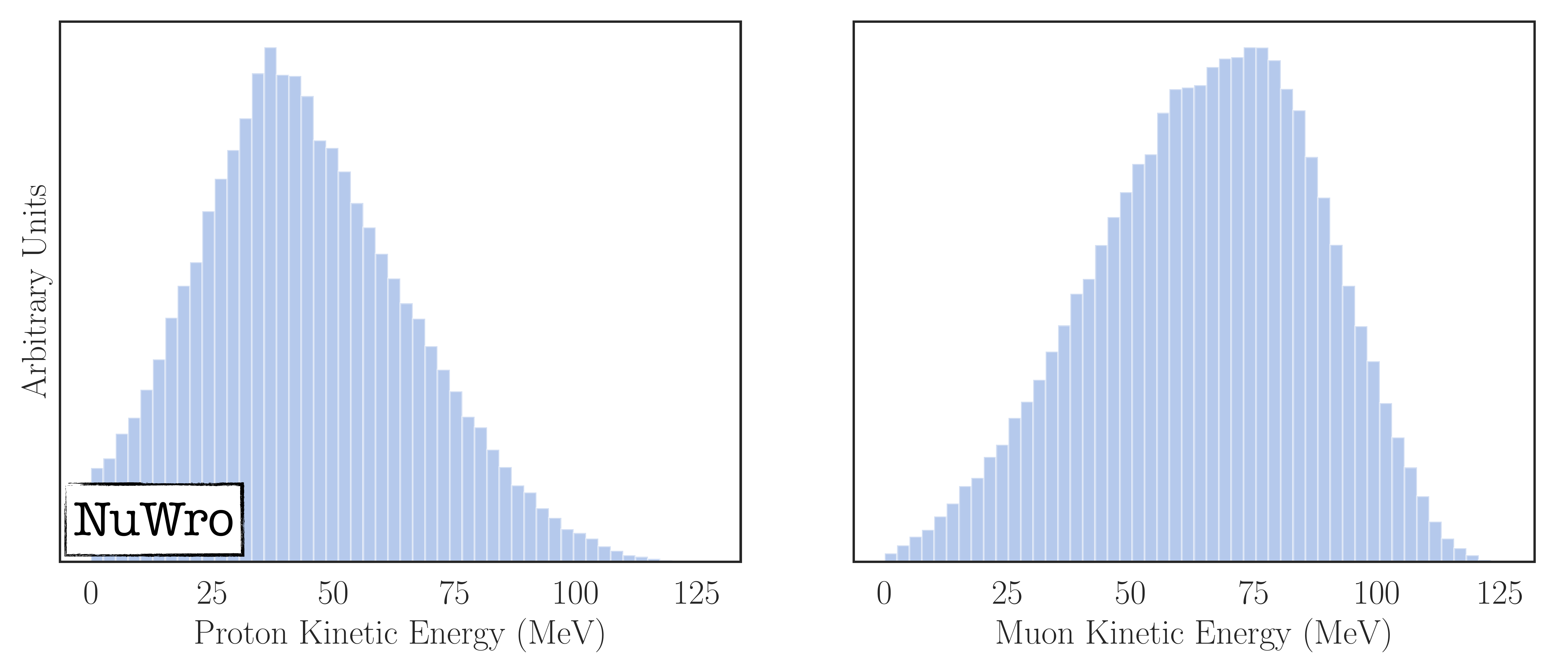}
  \cprotect\caption{\label{fig:KE} \textcolor{black}{NuWro \cite{Golan:2012rfa} generated proton and muon kinetic energies for the 1 proton + 1 muon = 2 total particles case. Generally, if we do not enforce a 2 particle cap, 13\% of the CC events are multi-proton at generator level. 76\% are single proton. 10\% are without protons. }}
\end{figure}
%=============================================================================

\subsection{Event Generation}

\par
We use \verb+NuWro+ \cite{Golan:2012rfa} to simulate neutrino-nucleus scattering events because it allows us to model the nuclear 
response using a spectral function to simulate the 
nucleus~\cite{Ankowski:2007wr}, 
rather than the Fermi Gas model .
Final state interactions are modeled using an \mbox{intra-nuclear} cascade (INC)~\cite{Golan:2012wx}.
\textcolor{black}{At 236 \mev, the \texttt NuWro neutrino 
event generator 
predicts a 4\% MEC (meson exchange current) contribution, a 32\% NCQE (neutral current quasi-elastic) contribution, and 
a 64\% CCQE (charged current quasi-elastic)  contribution
to the neutrino-argon scattering cross section, with a negligible contribution for all other processes (pions are produced 0.04\% of the time). However, neutral current interactions do not eject muons. We do not include NC in our analysis because we expect excellent muon identification in DUNE \textcolor{black}{and hence 
very few NC events in which a muon is identified.
This expectation is motivated by the success of the dE/dx vs.~residual range method at ProtoDUNE-SP (as shown  
in~\cite{ProtoDUNESP_Performance}).}}
\textcolor{black}{At $236 \mev$, neutrino charged-current interactions with nucleons are mostly quasi-elastic (CCQE), $\nu_{\ell} + n \rightarrow \ell^- + p^+$.} 
\textcolor{black}{Fig.~\ref{fig:KE} shows the expected distribution, generated by \textcolor{black}{\texttt NuWro}, of the kinetic energies of the muons and protons}
\textcolor{black}{produced by charged current interactions 
of a 236 MeV $\nu_\mu$.}

Thus, we are interested in charged-current quasi-elastic (CCQE)  
$\nu_\mu + \Ar{40}$ interactions. \textcolor{black}{We simulate CCQE
signal events  - $236 \mev$ neutrinos arriving from the direction of the Sun - and
background events (atmospheric neutrino events, assumed to be isotropic) 
in \mbox {\texttt NuWro.}} We do not consider non-DM KDAR in the Sun as a background. True, cosmic rays impinge on the Sun and produce KDAR but this contribution is negligible \cite{Rott:2015nma}.

\begin{figure}[hbt!]
  \centering
  \includegraphics[width= 1.0 \textwidth]{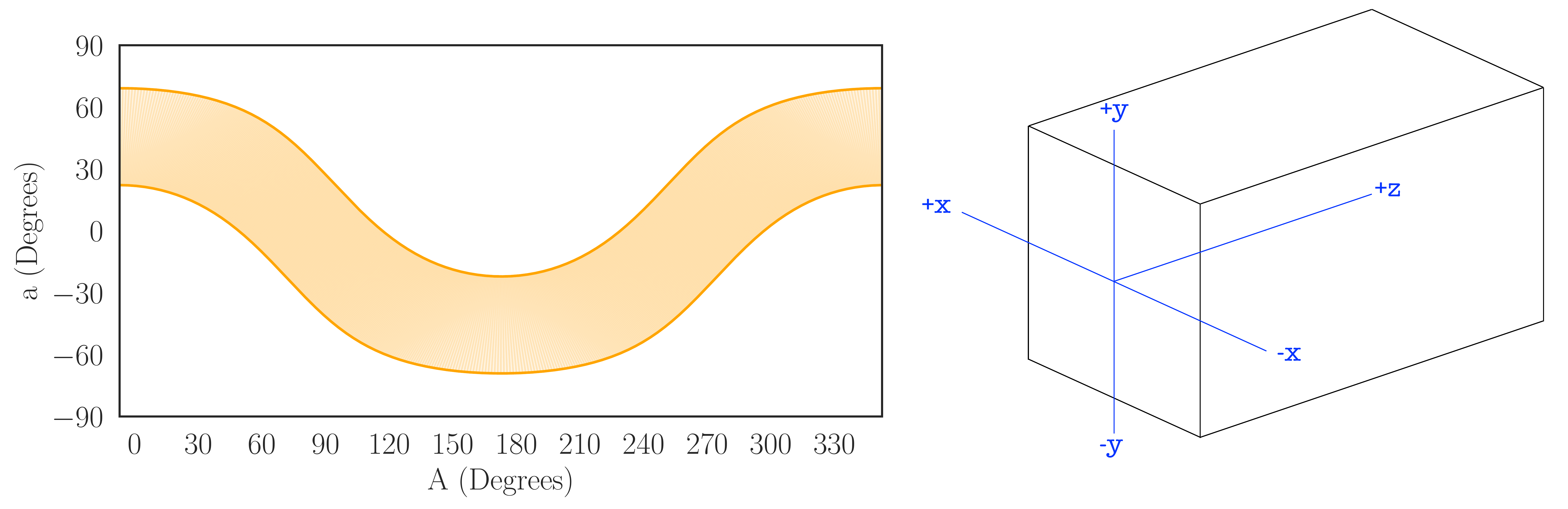}
  \cprotect\caption{\label{fig:sun_dir} \textcolor{black} {Solar directions (a = altitude, A = azimuth) seen at DUNE. The band is the finite angular coverage of the Sun. Azimuth winds clockwise from $\hat{x}$ to $\hat{z}$ in the detector frame and altitude goes up from the xz plane to $\hat{y}$. Note, these angles are often defined respect to the cardinal directions rather than the detector.}}
\end{figure}

 \textcolor{black}{For signal events, the neutrino is assumed to arrive from the direction of the Sun, but at a 
randomized time (which determines the orientation of the Sun with respect to the detector).
For an atmospheric neutrino background event, the orientation of the neutrino with respect 
to the detector is randomized. The distribution of signal event directions 
relative to the detector are show in 
Fig.~\ref{fig:sun_dir}.  In particular, and unlike 
atmospheric neutrinos, neutrinos arriving from the 
Sun cannot have an arbitrary orientation 
with respect to the 
detector wires, but must instead arrive from directions 
within the yellow band of Fig.~\ref{fig:sun_dir}.}

%=============================================================================
%=============================================================================

\subsection{Event Simulation and Reconstruction}

 \textcolor{black}{In each event, the particles generated with NuWro serve as input for LArSoft \cite{LArSoft}, which propagates the particles through argon (using GEANT4 \cite{GEANT4}) and simulates the detector response to the drifted ionization electrons. LArSoft also searches the simulated TPC wire waveforms for regions of interest and deconvolves and fits them to a Gaussian. These cleaned up ``hits" are 2D (each plane of wires is an image of ticks vs. wire) and are shown in Fig 1. Finally, Pandora \cite{Pandora}, a pattern recognition software kit, maps the 2D hits from the 3 wire plane projections to 3D and then clusters the 3D positions into tracks and showers.}

\subsection{Energy and Angular Resolution}
\begin{figure}[hbt!]
  \centering
  \includegraphics[width= 1.0 \textwidth]{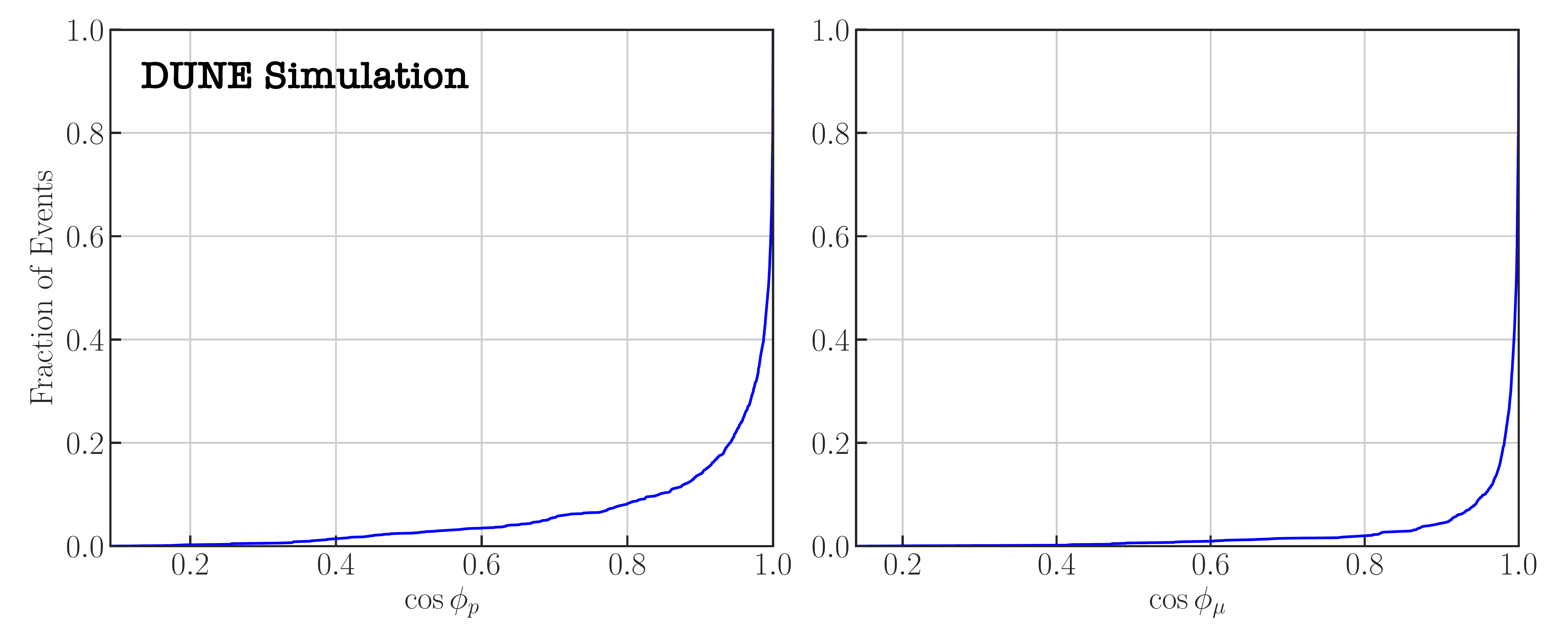}
  \cprotect\caption{\label{fig:angular_res} \textcolor{black}{ Cumulative distribution functions of the angular difference between the true and reconstructed track directions. $\phi_{p}$ is the proton angular difference and $\phi_{\mu}$ is the muon angular difference.}}
\end{figure}

\begin{figure}[hbt!]
  \centering
  \includegraphics[width= 1.0 \textwidth]{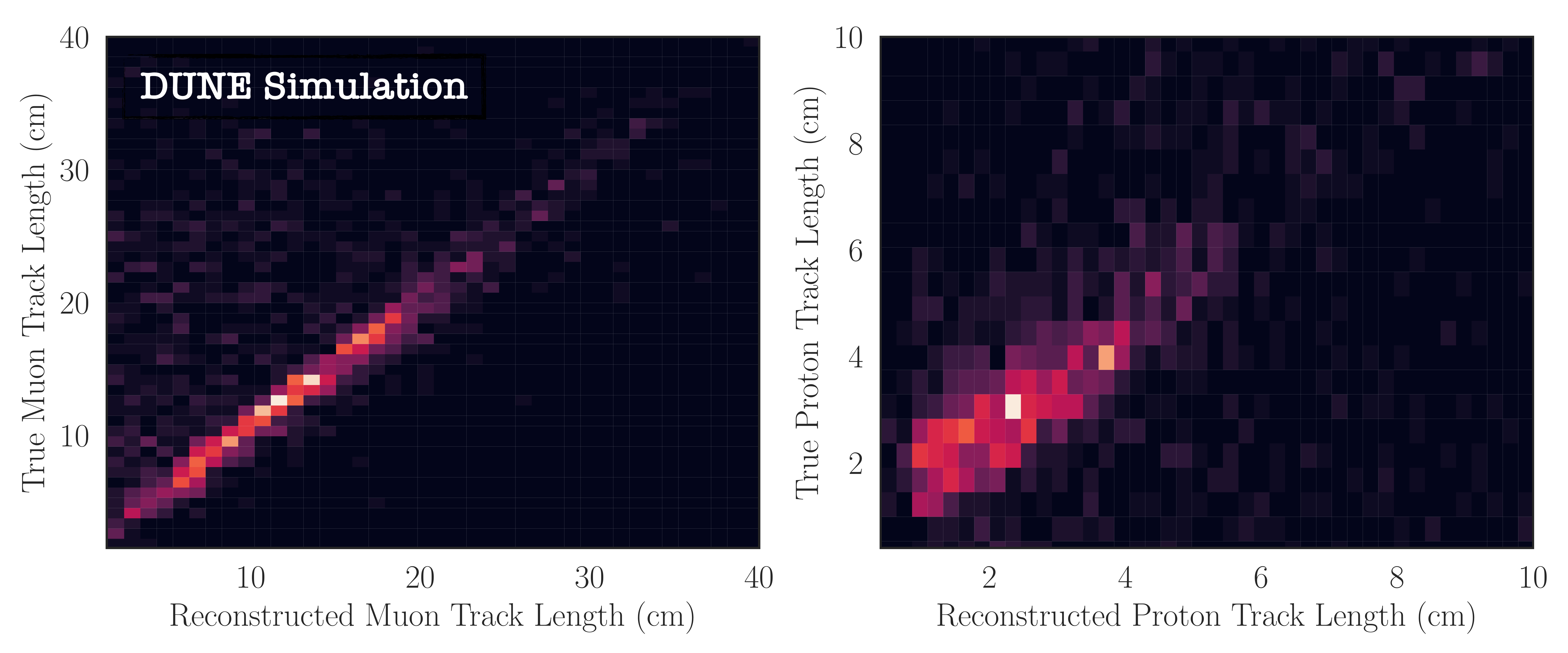}
  \cprotect\caption{\label{fig:lengths_comparison} \textcolor{black} { Comparison of the true (GEANT4) and the reconstructed track lengths for the muon (left) and proton (right). This analysis uses the track lengths to infer the proton and muon momenta.}}
\end{figure}

We can estimate the angular resolution with which proton 
and muon tracks can be reconstructed by comparing the 
direction of the outgoing particle at the event 
generator level to the direction of the fully reconstructed 
tracks.  We find that roughly $50\%$ of tracks are 
reconstructed to within $5^\circ$ of the true particle 
direction (Fig.~\ref{fig:angular_res}).  \textcolor{black}{Furthermore, we infer the particle momenta via ``range'' (track length). Fig.~\ref{fig:lengths_comparison} compares the true (GEANT4) and the reconstructed track lengths and gives us faith in this method. The true track length is the distance over which GEANT4 propagates the particle before it stops or decays, while the reconstructed track length is based on the hits generated by the ions created by this particle.}

%=============================================================================
%=============================================================================

%=============================================================================
%=============================================================================

\textcolor{black}{The charge read out on the LArTPC wires can be mapped to the kinetic energy of the culprit particle which caused the ionization. For events in 
which a proton and muon track are identified,  we can 
measure the proton
and muon energies, including the particle rest mass and the kinetic energy.}  

\begin{figure}[hbt!]
  \centering
  \includegraphics[width= 0.9 \textwidth]{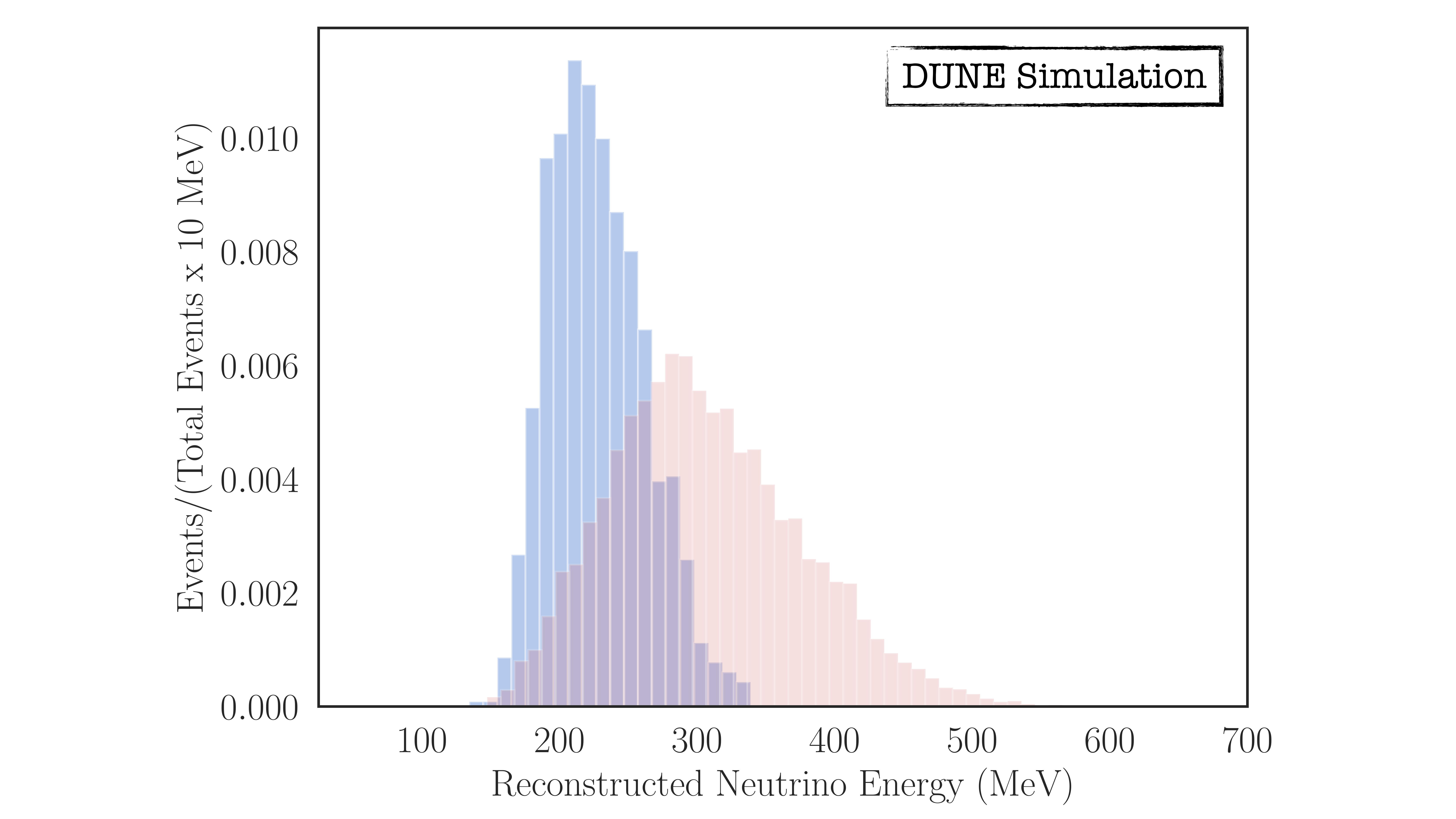}
  \cprotect\caption{\label{fig:e_res} \textcolor{black} {Distribution of the reconstructed neutrino energies. The blue histogram shows events with a true energy of 236 MeV, while the red histogram corresponds to the atmospheric background in the 150-400 MeV range. The area of both distributions is normalized to 1.} The signal (blue) has a standard deviation of $\pm 30 \mev$. This informs our choice of the relevant background energy - true energies between 150 and 400 MeV - a range 3 times larger than the reconstructed signal energy  resolution.} 
\end{figure}

We reconstruct the $\nu_\mu$ energy using the expression 
\bea
\label{eqn:e_eqn}
E_{\nu_{\mu}}^{\mathrm{recon}} \equiv E_{p} + E_{\mu} + (m_{\mathrm{Ar}}^{39} - m_{\mathrm{Ar}}^{40}).  
\eea
\textcolor{black}{In Fig.~\ref{fig:e_res}, we plot the 
distribution of reconstructed neutrino energies for events 
in which a 236 MeV $\nu_\mu$ charged-current interaction 
is simulated in NuWro.  The reconstructed neutrino energies 
are well clustered around the true energy of 236 MeV, with 
a variance of 30 MeV.} 
Eq.~\ref{eqn:e_eqn} does not include the 
kinetic energy of the remnant 
$\Ar{39}$.  Although the maximum momentum transfer to the nucleus is 
${\cal O}(200) \mev$, the maximum recoil energy is ${\cal O}(1) \mev$, \textcolor{black}{which is negligible compared to the  30 MeV energy resolution. %(Fig.~\ref{fig:e_res}).
}

%=============================================================================
%=============================================================================

\subsection{\textcolor{black}{Event Selection}}

\par
The atmospheric neutrinos are taken to have energies between 150\mev and 400\mev, \textcolor{black}{with an angle-averaged energy spectrum\footnote{\textcolor{black}{Besides an angle-averaged spectrum, \cite{Homestake} provides direction-dependent fluxes binned in the cosine of the zenith angle, Z,  and azimuth, $\phi$. The fractional variance of the 
direction-dependent fluxes compared to the angle-averaged flux decreases with energy for the energies relevant to this study. At $236 \, (600) \mev$, it is 0.34 \, (0.19). In using the average, the maximum overestimate at 236 MeV is a factor of 3.7. This happens between (-0.8,-0.9) in $\cos(Z)$ and between (90,120) degrees in $\phi$. The maximum underestimate is by a factor of 2.2. This happens between (0, 0.1) in $\cos(Z)$ and between (270, 300) degrees in $\phi$.}} calculated for Homestake at the solar minimum~\cite{Homestake}. We choose this background energy range in order to encompass 3 standard deviations of the reconstructed signal energy.  We are justified in ignoring 
atmospheric neutrinos whose true energies lie outside this
range, since they can be well distinguished from the 
signal by reconstruction of the neutrino energy.} 

\par
\verb+NuWro+ reports the neutrino-nucleus 
CCQE cross section; for signal events it reports the 
cross section at $E_{\nu_\mu} = 236~\mev$, and for 
atmospheric neutrinos it reports the average cross section
weighted by the neutrino energy spectrum
\textcolor{black}{between  
150\mev and 400\mev}. 
\par
These cross sections are
\bea
\sigma_{\nu_\mu-\Ar{40}}^{sig} &=& 
2.6 \times 10^{-38}~\cm^2 ,
\nonumber\\
\sigma_{\nu_\mu-\Ar{40}}^{bgd.} &=& 
2.8 \times 10^{-38}~\cm^2 .
\eea
\textcolor{black}{In simulating the CC cross section, we only have events with produced muons, and with neutrinos in the aforementioned energy range. The CCQE cross section is weighted and averaged only over this range.}
We have not simulated neutral current events, because such events do not 
produce a muon.

\par
As an initial event selection cut, we consider events in which exactly 
two tracks are reconstructed, 
that of a muon and a proton.
Although it is expected that DUNE will have excellent particle identification, for simplicity, 
we only require that Pandora identify exactly two tracks, 
and we assume that the longer track is a 
muon while the shorter track is a proton. \textcolor{black}{ At 236 MeV, GEANT4 predicts this to be the case 93\% of the time. Out of these 93\%, 97\% are correctly reconstructed as the longer track.}
Also, a small number of events 
passing the cuts contain additional ejected nucleons at the event generator 
level, but for which only one nucleon track was reconstructed.

\textcolor{black}{The requirement that we reconstruct the interaction with an interaction point within the fiducial 
volume justifies our assumption that the dominant background 
arises from atmospheric neutrinos.  There are a variety of 
other cosmogenic backgrounds at DUNE, but these 
backgrounds are 
unlikely to 
produce an identified  
muon track which is reconstructed to begin 
within the detector.  In other words, we have assumed 
that the analysis is based on a fiducial volume chosen 
such that the rate of such backgrounds is negligible.}

\subsection{Neutrino directionality}
Since the momentum transfer to $\Ar{39}$ is non-negligible, one cannot use 
$\vec{p}_\mu$ and $\vec{p}_p$ to reconstruct the direction 
of the incoming neutrino.\footnote{Note, for higher energy neutrinos, the momentum 
transfer to the remnant nucleus is negligible compared to the energy of neutrino, 
in which case the momentum of the charged lepton and of the hadronic ejecta is 
sufficient to reconstruct the neutrino direction effectively.  These techniques 
were used in~\cite{Rott:2019stu}.}
Instead we note that, given a hypothesis for the direction of the incoming neutrino, one 
can use momentum conservation to determine the momentum transfer to the remnant nucleus.  \textcolor{black} {We define the kinematic variable 
\bea
\vec{p}_{\Ar{39}} &\equiv& 
(236~\mev) \hat p_\odot - \vec{p}_\mu - \vec{p}_p ,
\eea
where $\hat p_{\odot}$ is a unit vector pointing from the Sun 
to the detector.}  If the incoming $236 \mev$ neutrino were 
actually arriving from the Sun, then
$\vec{p}_{\Ar{39}} $ would be the 
\textcolor{black}{reconstructed}
momentum of the 
remnant nucleus.

\begin{figure}[hbt!]
  \centering
  \includegraphics[width= 1.0 \textwidth]{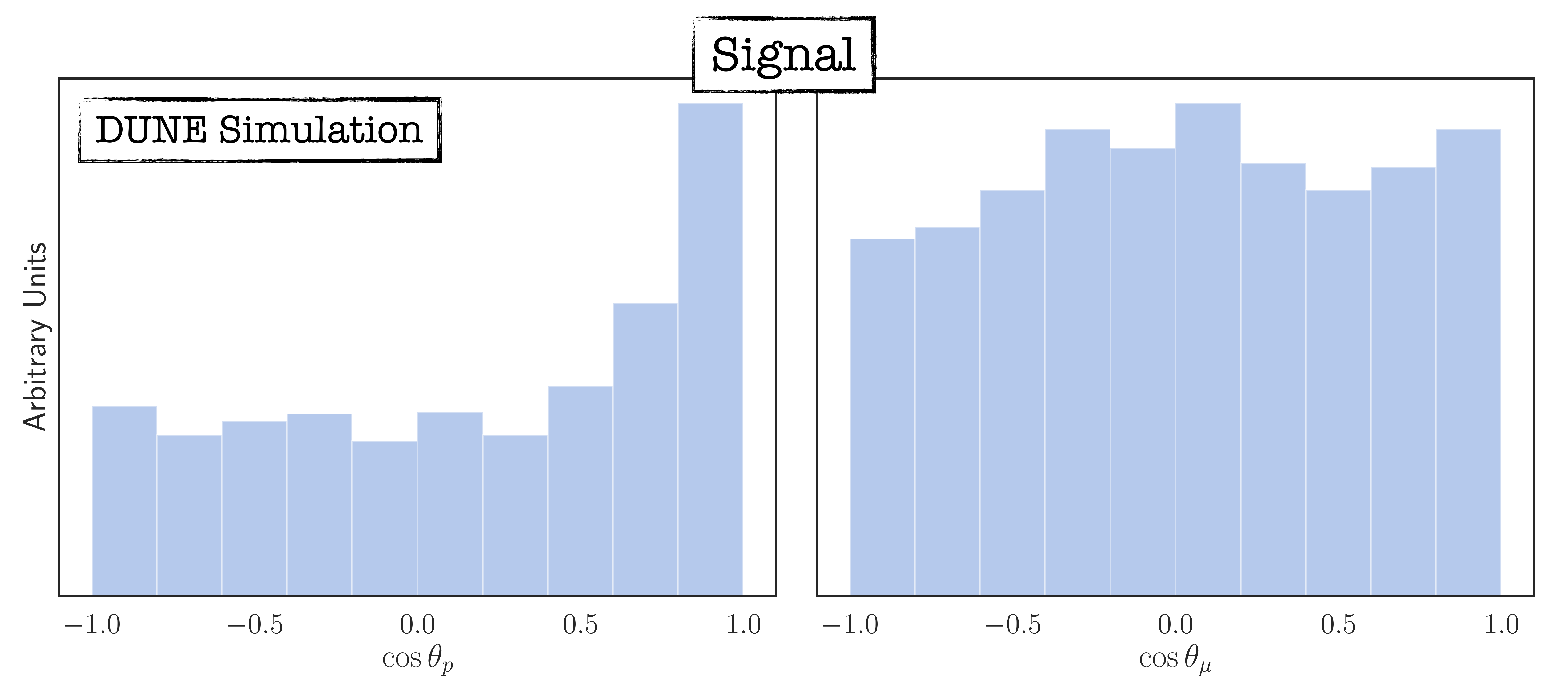}
  \cprotect\caption{\label{fig:reco_angles} \textcolor{black} { Signal distribution of the reconstructed proton and muon angles respect to the incoming neutrino}. The proton tends to fly out forward.}
\end{figure}

As noted in~\cite{Rott:2016mzs}, the ejected proton tends to emerge preferentially in the 
forward direction.  As such, the angle $\theta_p$ between the proton and the direction from the 
Sun, defined by \mbox{$\cos \theta_p = (\hat p_\odot \cdot \vec{p}_p) /|\vec{p}_p|$}, is one of 
the kinematic variables upon which we will impose cuts (Fig.~\ref{fig:reco_angles}).

%=============================================================================
%=============================================================================

%=============================================================================
\begin{figure}[hbt!]
  \centering
  \includegraphics[width= 1.0 \textwidth]{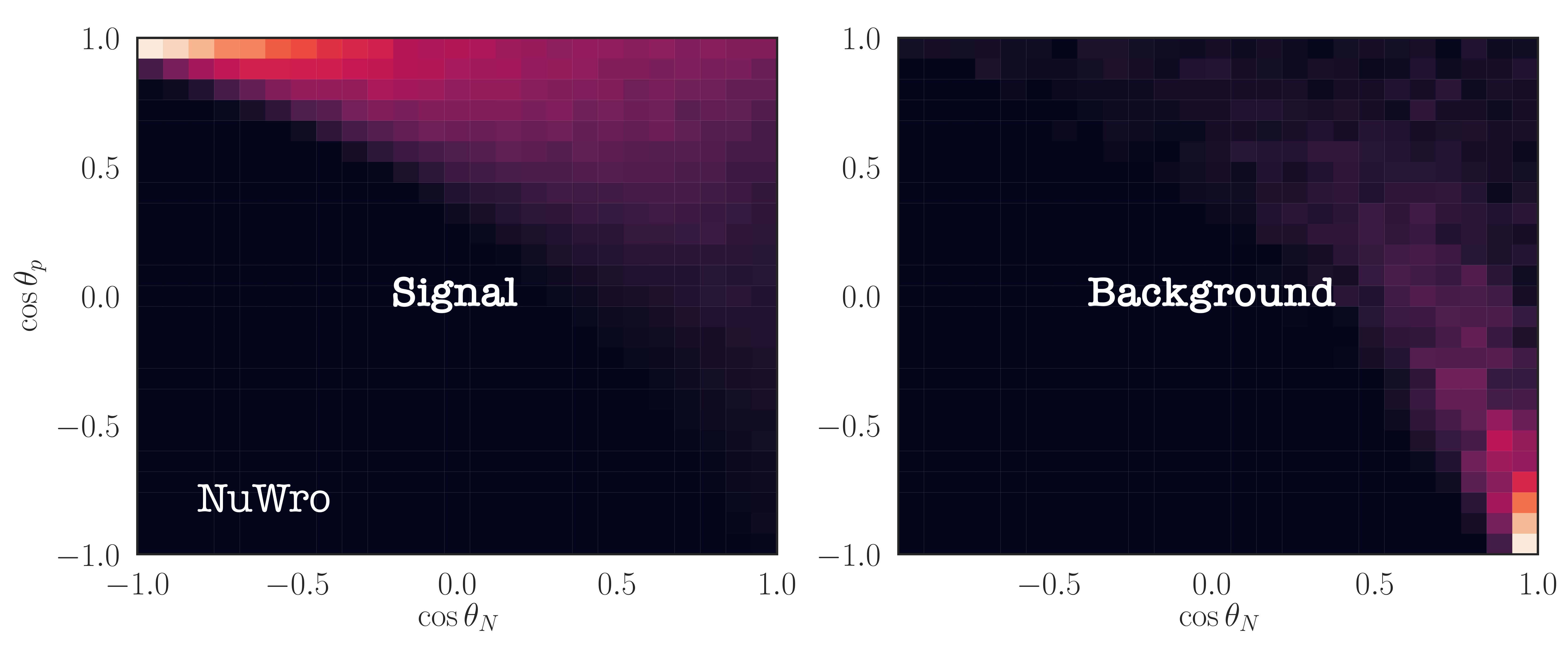}
  \caption{\label{fig:mc_dircut} \textcolor{black}{2D histograms of the proton and the remnant nucleus angles respect to the incoming neutrino at the generator level (not put through the detector simulation and reconstruction). These events passed our aforementioned topology and energy selection. Note, $\cos \theta_N$ and $\cos \theta_p$ are always well-defined. Signal/background is on the left/right. For the atmospheric background, assuming that the incoming neutrino points to the Sun, rather than isotropically, violates momentum conservation and leads to an incorrect nuclear recoil and a distinct angular distribution.}}
\end{figure}
%=============================================================================

%=============================================================================
\begin{figure}[hbt!]
  \centering
  \includegraphics[width= 1.0 \textwidth]{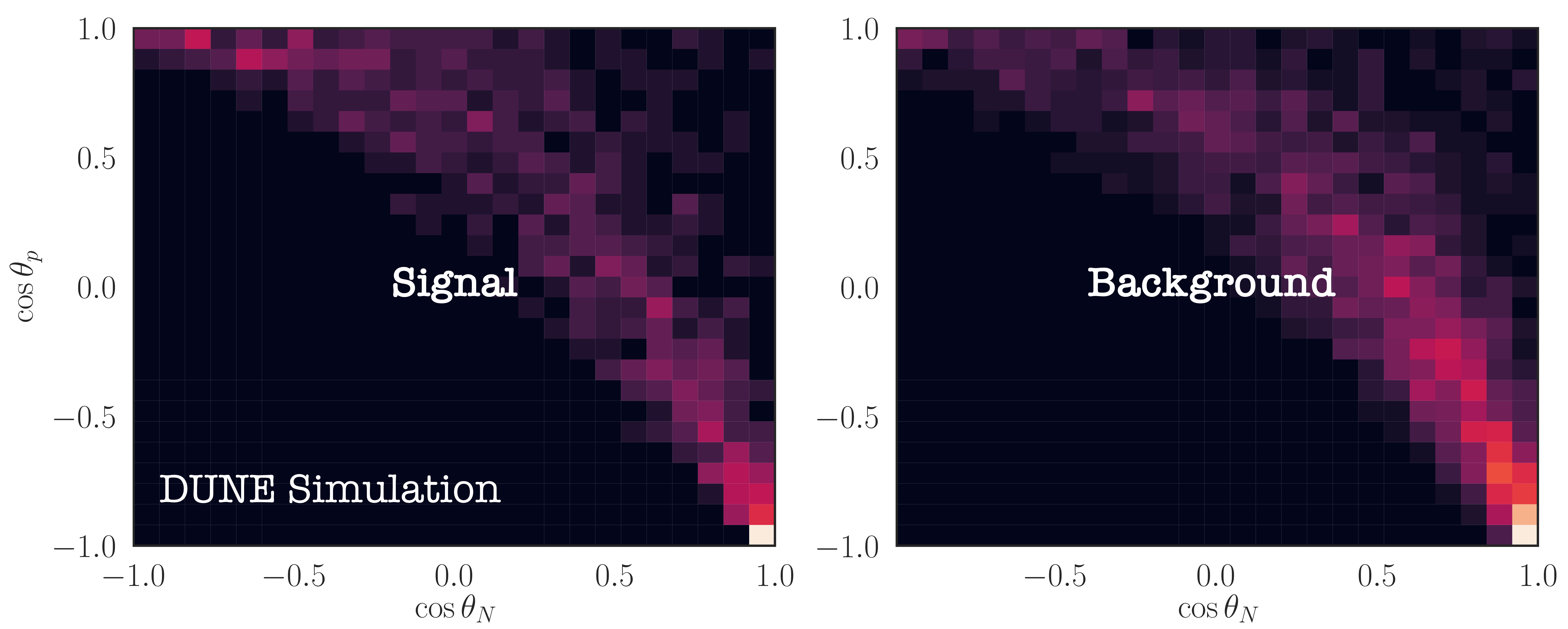}
  \caption{\label{fig:dircut} Reconstructed \textcolor{black}{2D histograms} of
  \textcolor{black}{$\cos \theta_p$ and $\cos \theta_N$}.
  These events have gone through the detector simulation and reconstruction and passed the energy cut. Comparing to Fig.~\ref{fig:mc_dircut}, the discrimination between the signal (left) and background (right) is much reduced due to poor reconstruction of back to back tracks. }
\end{figure}
%=============================================================================

We also find that a useful kinematic variable is \mbox{$\theta_N$, defined by 
$\cos \theta_N \equiv \hat p_\odot \cdot \vec{p}_{\Ar{39}} / |\vec{p}_{\Ar{39}}|$.}
\textcolor{black}{If the neutrino does indeed arrive from the direction 
of the Sun with an energy of 236 MeV, 
then $\theta_N$ would evaluate to the 
angle between 
the reconstructed remnant nucleus momentum and the direction of the Sun.}
\textcolor{black}{We plot a generator level (reconstruction level) 2D histogram of $\cos \theta_p$ vs. $\cos \theta_N$ in Fig.~\ref{fig:mc_dircut} (Fig.~\ref{fig:dircut}).}

Unsurprisingly, both signal and background 
distributions contain events in which $\cos \theta_N$ is 
close to 1, since the definition of 
$\vec{p}_{\Ar{39}}$ biases it in the forward 
direction.  Perhaps more surprisingly, the signal 
distribution contains a significant population of events 
in which $\cos \theta_p \sim 1$, while 
$\cos \theta_N \sim -1$.  There is no similar population of 
events in the background distribution, implying that a good 
way to reject background is to select events in which the 
proton is ejected in the direction away from the Sun, while 
\textcolor{black}{$\vec{p}_{\Ar{39}}$}
points back to 
the Sun.

After 
reconstruction (Fig.~\ref{fig:dircut}), the discrimination between signal and background is poorer. \textcolor{black}{Although the angular distribution for the charged lepton is isotropic, it is nevertheless correlated with that of the proton;} for events where the ejected proton and recoiling nucleus are 
(anti-)collinear with the neutrino, the charged lepton track also lies upon the 
same line.  In this case, the reconstruction algorithm may be unable to distinguish 
the proton and charged leptons tracks, leading to an event reconstructed with just a 
single track, which would be rejected by the event selection cuts. However, we will see that the shift in 
dark matter sensitivity due to this reconstruction failure is less 
than $\mathcal{O}(10)$. 

\begin{figure}[hbt!]
  \centering
  \includegraphics[width= 0.7 \textwidth]{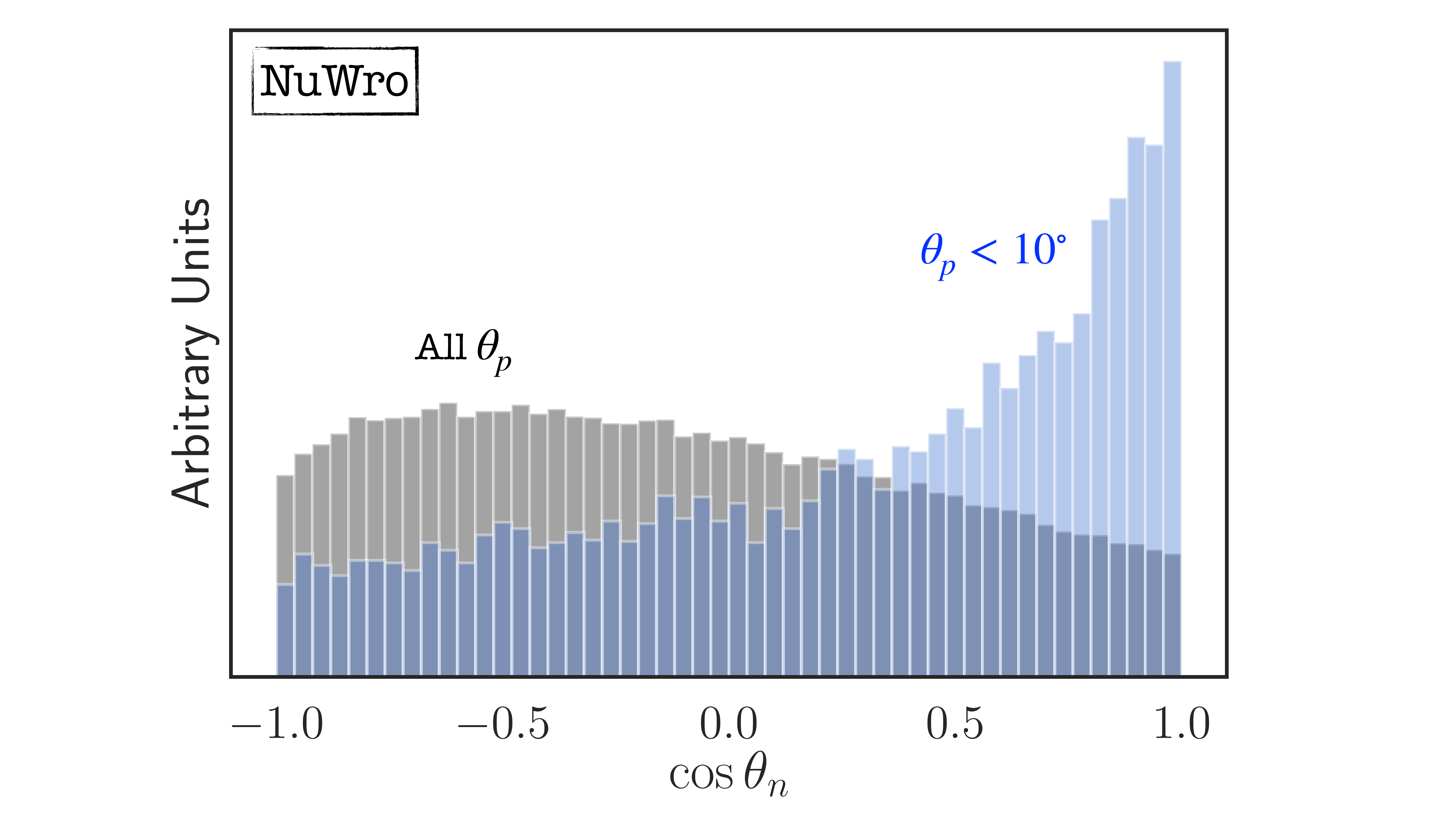}
  \cprotect\caption{\label{fig:nucleon} \textcolor{black}{NuWro generated (normalized) distribution of the angles of the struck nucleons %which would be scattered 
  - before scattering %they were scattered 
  - with respect to the incoming signal neutrino arriving from the Sun. For the grey histogram, only the event selection cuts are imposed. The blue histogram is for cases with forward ejected protons ($\theta_p < 10^{\circ}$). This figure shows that such protons are correlated with nucleons which were forward-going before the interaction.}}
\end{figure}

It may seem counterintuitive that the remnant nucleus should 
be backscattered in CCQE events.  But an examination of the 
corresponding events at generator level provides an 
explanation; in the majority of events in which the proton 
is forward-directed and the remnant nucleus is 
backward-directed, the nucleon struck by the neutrino had 
an initial momentum in the direction away from the Sun (Fig.~\ref{fig:nucleon}). When the struck nucleon is already moving away from the Sun, the 
outgoing nucleon is also typically very forward-directed, 
while the remaining nucleons have a net momentum in 
the opposite 
direction, leading to a backward directed remnant nucleus.

%=============================================================================
%=============================================================================

%=============================================================================
\begin{figure}[hbt!]
  \centering
  \includegraphics[width= 1.0 \textwidth]{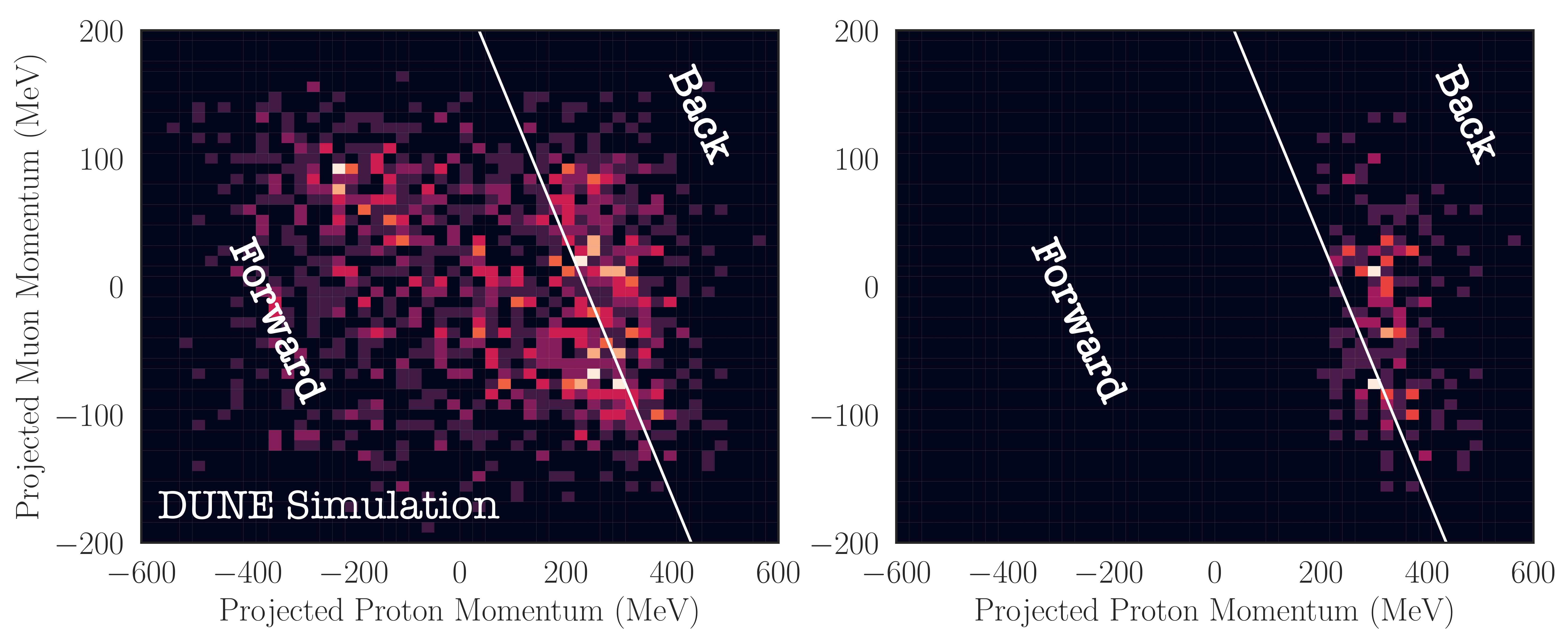}
    \caption{\label{fig:nuc_recoil} \textcolor{black}{Signal 2D histograms} of the reconstructed muon and proton momenta projected along the incoming neutrino direction. The left plot has only event selection cuts, 
  whereas the right plot only includes such events with very forward protons ( $\cos \theta_{p} < 30^{\circ}$ ). The white line 
  ($( \vec p_{\nu} -  \vec p_{p}  - \vec p_{\mu} ) \cdot \hat p_{\nu} =0 $) separates forward/backscattered remnant nuclei. \textcolor{black} {Momentum conservation means that 
  $\vec p_{\Ar{39}}  =  \vec p_{\nu}- \vec p_{p}  - \vec p_{\mu} $.}
%  $( \vec p_{N}  =  \vec p_{\nu}- \vec p_{p}  - \vec p_{\mu} ) \cdot \hat p_{\nu}$. 
  Hence, the nucleus backscatters when $\vec p_{\Ar{39}} \cdot \hat p_{\nu}  < 0$ (to the right of the line). This figure emphasizes that the remnant nucleus tends to be backscattered if the proton is very forward scattered.  }

\end{figure}
%=============================================================================
To illustrate this point, 
we plot the distribution of signal events in the 
$(\vec{p}_p\cdot \hat p_\odot, \vec{p}_\mu \cdot \hat p_{\odot})$ plane (Figs.~\ref{fig:nuc_recoil}).  The left panel is the distribution 
of all signal events passing event selection cuts, while 
the right panel is the distribution of such events for 
which $\cos \theta_p < 30^\circ$.  \textcolor{black} {In both panels, the 
white diagonal line indicates $(\vec{p}_p+\vec{p}_\mu)
\cdot \hat p_\odot =236~\mev$; events to the right of 
this line have 
\textcolor{black}{$\cos \theta_N <0$,}
while events to the left 
have 
\textcolor{black}{$\cos \theta_N > 0$,}.}

\textcolor{black}{We define the signal efficiency 
$\eta_S^\mu$ to be the fraction of 
signal neutrino events which pass the event selection 
cuts as well as the the energy and directionality cuts we 
impose.  Similarly, we define the background efficiency 
$\eta_B^\mu$ to be the fraction of 
atmospheric neutrino events with a neutrino energy between  
$E^{bgd}_{min} = 150\mev$ and $E^{bgd}_{max} = 400\mev$ 
which pass these cuts.
Only a negligible fraction of atmospheric neutrinos outside the range $150-400~\mev$ pass the cuts.}

 Motivated by the reconstructed energy resolution of the signal events, we  impose an energy cut by selecting 
 only events with reconstructed neutrino energy in the range $236\pm 30 \mev$. Also, since protons often fly out forward, we require them to lie within an angular cone centered on the direction pointing from the Sun.
 A similar approach for the leptons is fruitless. At such energies, their ejection is largely isotropic. \textcolor{black}{Finally, we impose cuts on 
 $\cos \theta_N$.}

Various cuts and their effect on DUNE's sensitivity to a 236 MeV flux of $\nu_{\mu}$ emanating from the Sun are listed in Table~\ref{tab:Cuts}.

\section{Solar KDAR $\nu_{\mu}$ Flux}
We will first determine the number of background atmospheric 
neutrino events which are expected to pass our cuts over 
a given exposure of DUNE.  
\bea
\label{eqn:NB}
N_B^{\mu} &=& \eta_B^{\mu} \int_{E_{min}^{bgd}}^{E_{max}^{bgd}} dE_\nu~ d\Omega ~ \frac{d^2 \Phi_B^{\mu}}{dE_\nu d\Omega} \times \left( \bar A_{\rm eff}^{(\mu)}  T \right) ,
\eea
\textcolor{black}{where $\eta_B^{\mu}$, $E_{min}^{bgd}$ and $E_{max}^{bgd}$ 
are defined as in the previous section. 
$d^2 \Phi^\mu / dE_\nu d\Omega$ is the differential flux 
of atmospheric $\nu_\mu$, and $T$ is the exposure time. 
The effective area of  DUNE effective  is the 
product of the neutrino-nucleus scattering cross section 
with the number of nuclei in the fiducial volume.  
We take DUNE's effective area to atmospheric $\nu_\mu$, 
$\bar A_{\rm eff}^{(\mu)}$, to be given by }
\bea
\label{eqn:Aeff}
\bar A_{\rm eff}^{(\mu)} &=& (6.0 \times 10^{-10}\m^2 )
\left( \frac{\sigma_{\nu \text{-Ar}}^{(\mu)bgd.} }{10^{-38}\cm^2 } \right) \left( \frac{M_{\rm target}}{40 \kT} \right) ,
\eea
\textcolor{black}{where $\sigma_{\nu \text{-Ar}}^{(\mu)bgd.}$ is the $\nu_\mu$-Ar charged-current scattering 
cross section, 
weighted by the atmospheric neutrino spectrum in the 
energy range $\left(E_{min}^{bgd}, E_{max}^{bgd} \right)$, as described 
in Section 2.  Combining \ref{eqn:NB} and \ref{eqn:Aeff} gives }
%yielding
\bea
N_B^{(\mu)} 
 &=& \eta_B^{\mu}  (2.39  )
\left( \frac{\sigma_{\nu \text{-Ar}}^{(\mu)bgd.} }{10^{-38}\cm^2 } \right) \left( \frac{M_{\rm target} T}{400 \kT  \yr} \right)
\int_{E_{min}^{bgd}}^{E_{max}^{bgd}} dE_\nu ~ \frac{d^2 \Phi_B^{\mu}}{dE_\nu d\Omega} (\m^2 \s \sr).
\eea
Setting $E_{min}^{bgd} = 150 \mev$, $E_{max}^{bgd}=400 \mev$, and $\sigma_{\nu \text{-Ar}}^{(\mu)bgd.} = 2.80346 \times 10^{-38} \cm^2$, we can integrate the spectrum 
from~\cite{Homestake} calculated at Homestake at solar minimum, yielding 
\bea
N_B^{(\mu)} &=& \eta_B^{\mu} \times (6.67 \times 10^3) 
\times ({\rm exposure} / 400 \kT \yr).
\eea
Given the background acceptances $\eta_B^\mu$ listed in 
Table~\ref{tab:Cuts}, we can then determine the  
number of background events expected to pass the cuts, also 
listed in Table~\ref{tab:Cuts}.  

\textcolor{black}{
We assume that the number of signal and background events 
seen by DUNE will be drawn from  Poisson-distributions 
whose means are given by the expected number of signal 
and background events, denoted by $N_S^\mu$ and $N_B^\mu$, 
respectively.
To estimate the sensitivity of DUNE, we assume  a 
representative (``Asimov"~\cite{Cowan}) data set 
in which the number 
of observed neutrinos is taken to be the number of expected 
background neutrinos, rounded to the nearest integer (that 
is, $N_O^\mu = {\rm round}(N_B^\mu)$).  We denote by 
$N_S^{\mu,90}$ the number of expected signal events 
such that the likelihood of an experimental run 
observing a number of total events larger than 
${\rm round}(N_B^\mu)$ is 90\%.  A model for which the 
expected number of signal events satisfies 
$N_S^\mu > N_S^{\mu,90}$ lies in the region to 
which we estimate DUNE would be sensitive.
}

Given $N_S^{\mu, 90}$ and $\eta_S^\mu$, we can then 
straightforwardly determine $\Phi_{236\mev}$, the 
maximum flux of 236 MeV neutrinos emanating from the core 
of the Sun which would be allowed 
(at 90\% CL), given that 
%a 400 kT yr run of 
DUNE observed 
only a number of events consistent with atmospheric neutrino 
background.
\bea
\Phi_{236\mev} &=& \frac{N_S^{\mu, 90}}{\eta_S^\mu A_{\rm eff}^{(\mu)} (E_\nu) T} 
= 5.3 \m^{-2} \s^{-1} \frac{N_S^\mu}{\eta_S^\mu} 
\left( \frac{\sigma_{\nu \text{-Ar}}^{(\mu)}(E_\nu)}{10^{-38}\cm^2 } \right)^{-1}
\left(\frac{\text{exposure}}{400 \kT \yr}\right)^{-1},
\eea
where 
\bea
A_{\rm eff}^{(\mu)} (E_\nu) 
&=&  (6.0 \times 10^{-10}\m^2 )
\left( \frac{\sigma_{\nu \text{-Ar}}^{(\mu)}(E_\nu)}{10^{-38}\cm^2 } \right) \left( \frac{M_{\rm target}}{40 \kT} \right) ,
\label{eq:effArea}
\eea
and $\sigma_{\nu \text{-Ar}}^{(\mu)} (E_\nu =236 \mev) = 2.6 \times 10^{-38} \cm^2$.

$\Phi_{236 \mev}$ is our primary result, and represents the 
minimum flux of 236 MeV $\nu_\mu$ emanating from the core 
of the Sun to which DUNE would be sensitive with any given 
%a 400 kT yr 
exposure.  This result is independent of the 
the specific model of new physics which generates this 
excess flux of neutrinos, but is determined only by the 
efficiency with which 236 MeV neutrinos from the core of 
the Sun and atmospheric background neutrinos pass the cuts.

\begin{figure}[!htb]
  \centering
  \includegraphics[width= 0.9 \textwidth]{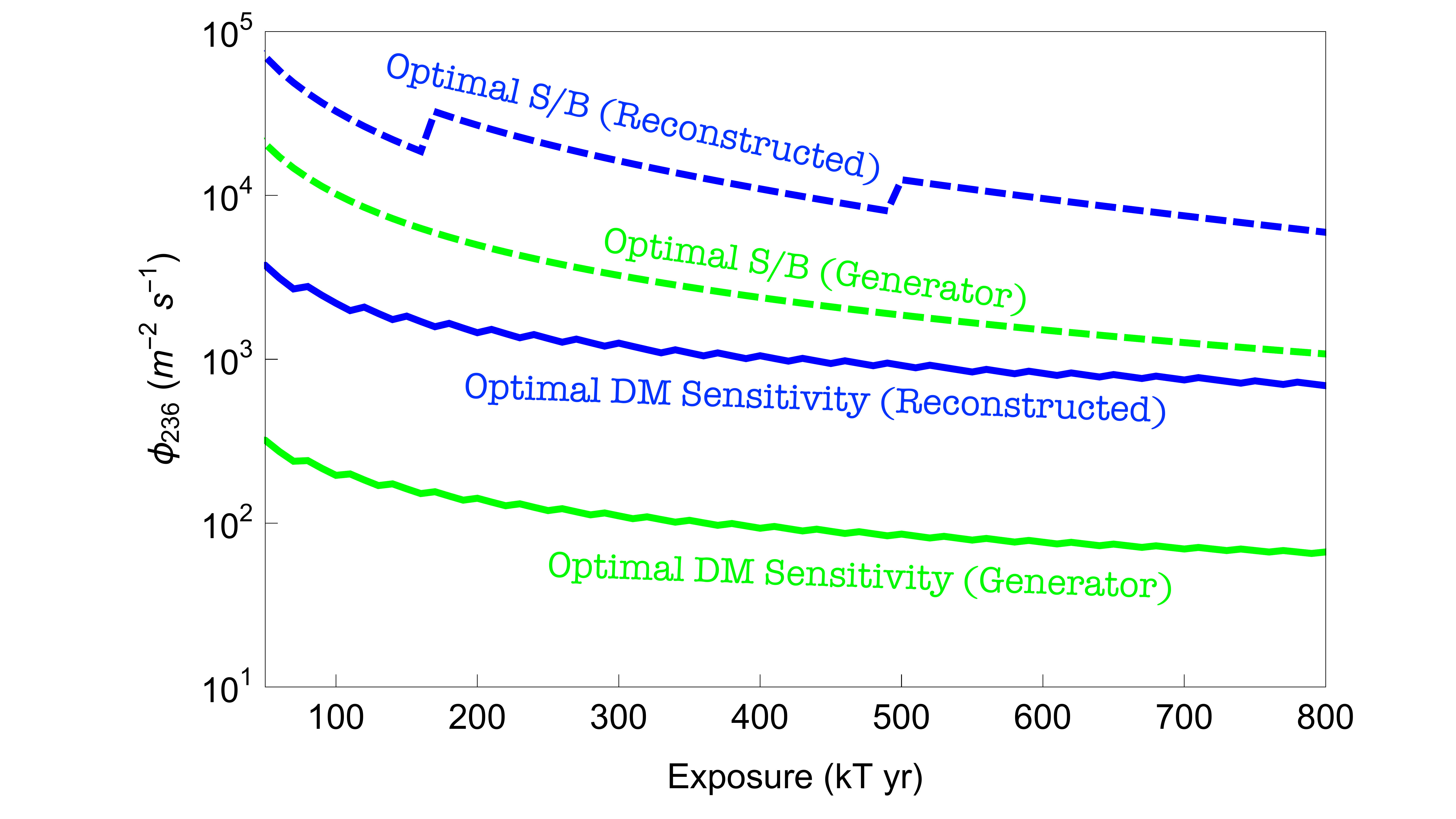}
  \caption{\label{fig:flux236} \textcolor{black}{ DUNE sensitivity to $\phi_{236 \mev}$, the flux of 236 MeV $\nu_{\mu}$'s from the Sun, independent of the new physics model that produces them. 
  The blue lines correspond to the reconstructed quantities in the first and second rows of Table \ref{tab:Cuts} and the green lines correspond to the generator level quantities in the third and fourth rows. The dashed lines are for maximum S/B and the filled lines are for maximum DM sensitivity. 
  The discontinuities are due to the limit of small numbers of events; noticeable when the number of observed events jumps by one. }}
\end{figure}

\textcolor{black}{
We plot $\Phi_{236 \mev}$ in 
Figure~\ref{fig:flux236}, as a function of the 
exposure,
\textcolor{black}{for several different choices of 
cuts (see Table \ref{tab:Cuts}).  In each case, the reconstructed neutrino energy is required to be in the range $236 \pm 30 \mev$. In one case, cuts on $\theta_N$ and $\theta_p$ are chosen to 
optimize signal significance (solid lines), while in 
the other case, these cuts are chosen to optimize 
the signal-to-background ratio (that is, $\eta_S/\eta_B$) (dashed lines).  
To illustrate the effect of possible improvements in 
track reconstruction, we also apply this analysis framework 
directly to the muon and proton tracks produced by 
the event generator; these curves are presented as green 
lines.  All four of the angular cut choices, along with 
their efficiencies, sensitivities, signal-to-background 
ratios, and number of expected signal and background 
events, are listed in Table~\ref{tab:Cuts}.}
For the 
cuts (applied to reconstructed events) which maximize 
the $S/B$, the sensitivity varies discontinuously.  This 
is because, in this case, the number of assumed events 
observed is small, and the jumps are where they vary 
discontinuously.}

%=============================================================================
%=============================================================================

\subsection{Application: Search for Inelastically Scattered Dark Matter}

To place this result in context, we consider a dark matter scenario which 
can be constrained by data from DUNE, but which would be difficult to constrain 
with direct detection experiments.  In particular, we consider the case of 
low-mass dark matter ($m_X \lesssim 10\gev$) which scatters inelastically with 
nuclei, with the emerging dark particle being $\delta = 50\kev$ heavier than the incoming 
dark matter particle.  In this case, dark matter inelastic scattering is kinematically inaccessible 
for detectors on Earth, because there is insufficient energy to produce the excited state.  
But because dark matter accelerates as it approaches the Sun, it may have sufficient 
kinetic energy to scatter inelastically against solar nuclei, leading to its gravitational 
capture~\cite{Gould:1987ir,Gould:1991hx,Garani:2017jcj,Nunez-Castineyra:2019odi}.

\textcolor{black}
{One example of a scenario in which inelastic scattering can dominate is the case in which the dark matter is charged under a spontaneously-broken $U(1)$ gauge symmetry.  
In this case, a dark matter vector current couples to the dark photon, which can be 
mediate dark matter-nucleon scattering.  The tree-level scattering process is 
necessarily inelastic, because the vector current for a single real particle vanishes.  
Elastic scattering is instead subleading, mediated either by multiple dark photon exchange 
or by other mediators with small couplings.  Although the size of this subleading elastic 
scattering cross section is model-dependent, it can be well below current direct detection 
sensitivity.}

\textcolor{black}
{After the initial inelastic scatter, the dark matter is gravitationally captured, and 
continues to orbit the Sun.  As the dark matter passes through the Sun many times, 
subsequent inelastic or elastic scatters result in an even greater loss of dark matter 
kinetic energy, until the particle settles in the core of the Sun~\cite{Nussinov:2009ft,Menon:2009qj}.  Once the dark matter has lost 
enough kinetic energy, inelastic scattering is no longer kinematically possible, but since the 
dark matter continues to pass through the Sun many times during the Sun's lifetime, even the subleading 
elastic scattering cross section can be sufficient to deplete the dark matter kinetic energy 
enough for it to settle in the core.
}

After gravitational capture, we assume  dark matter 
annihilation to first generation quarks,  with 
dark matter capture
and annihilation being in 
equilibrium.  
\textcolor{black}
{Even though dark matter annihilation produces only first-generation quarks, a substantial number of 
kaons are produced by subsequent fragmentation and 
hadronization processes.  
If the dark matter mass is $\gtrsim {\cal O}(5\gev)$, then the center of mass energy is 
large compared to the kaon mass, and  the up, down, and strange quarks can all be treated as 
light quarks.  }

We assume that dark matter scattering with nuclei is 
spin-independent and velocity-independent, with an equal coupling to protons 
and neutrons.  Because $\delta \ll m_X$, the 
dark matter-nucleon scattering matrix element is largely independent of $\delta$.  
The dependence of the dark matter-nucleus scattering cross section on $\delta$ 
arises from the final state phase space.  Thus, we will parameterize the dark matter 
model by $\sigma_0$, which is the total cross section for dark matter-nucleon scattering, 
extrapolated to $\delta = 0$.  From this quantity, the differential cross section 
for scattering against any nucleus at $\delta = 50\kev$ can be determined.

\textcolor{black}{In this scenario, the DM annihilation 
rate ($\Gamma_A$) is equal to one-half of the 
dark matter capture rate ($\Gamma_C$).  
The capture rate is directly proportional to $\sigma_0$, with 
$\Gamma_C = C_\delta (m_X) \times \sigma_0$.
The proportionality constant   
$C_\delta (m_X)$ is determined entirely by the dark 
matter mass, by solar 
physics, and the assumption that dark matter has a nominal 
Maxwell-Boltzmann velocity distribution with a density of 
$0.3 \gev/\cm^3$.  Relevant values for the 
$C_\delta (m_X)$ can be found in~\cite{Kumar:2012uh}. }

In this scenario, we can relate $\Phi_{236\mev}$ to 
$\sigma_0$, finding
\bea
\Phi_{236\mev} &=& 
\frac{(C_\delta (m_X) \times \sigma_0 / 2)
F^{\mu} }{4\pi r_{\oplus}^2 } \left( 0.64 \times \frac{2m_X}{m_K} r_{K}(m_X) \right), 
\nonumber\\
&=& (3.1 \times 10^4 \m^{-2} \s^{-1}) 
\left(\frac{C_{\delta} (m_X)}{10^{29} \pb^{-1} \s^{-1}} \right)
\left( \frac{\sigma_0}{\pb}\right)
\left(\frac{2m_X}{m_K} r_{K}(m_X) \right)
\eea
where $F^\mu = 0.27$ is the fraction of 236 MeV neutrinos 
which arrive at the detector as $\nu_\mu$, assuming a 
normal hierarchy.  
\textcolor{black}{While an experimental data analysis requires a full treatment of neutrino oscillations to obtain neutrino spectra and flavor ratios for specific times of detector operation, for this analysis it is sufficient to assume an annual averaged flavor ratio taken from~\cite{Lehnert:2007fv}}
\textcolor{black}{ (if one assumed an inverted hierarchy $F^\mu$ would increase by at most $25\%$).}
%~\cite{Lehnert:2007fv}.  
$r_\oplus = 1.5 \times 10^{11}\m$ is the 
distance from the Sun to the Earth, and 
$r_K (m_X)$ is the fraction of the center of mass energy 
of the dark matter initial state which is converted into 
stopped $K^+$ through dark matter annihilation, the 
hadronization and fragmentation of the outgoing particles, 
and the interactions of those particles with the dense 
solar medium (values for $r_K (m_X)$ can be found 
in~\cite{Rott:2015nma}).  
The factor $0.64$ is the branching fraction 
for $K^+$ decay to produce 
a monoenergetic 236\mev $\nu_\mu$. 
We can thus relate $\Phi_{236\mev}$ to a 90\% CL exclusion 
contour in the $(m_X, \sigma_0)$-plane.

\begin{figure}[hbt!]
  \centering
  \includegraphics[width= 0.9 \textwidth]{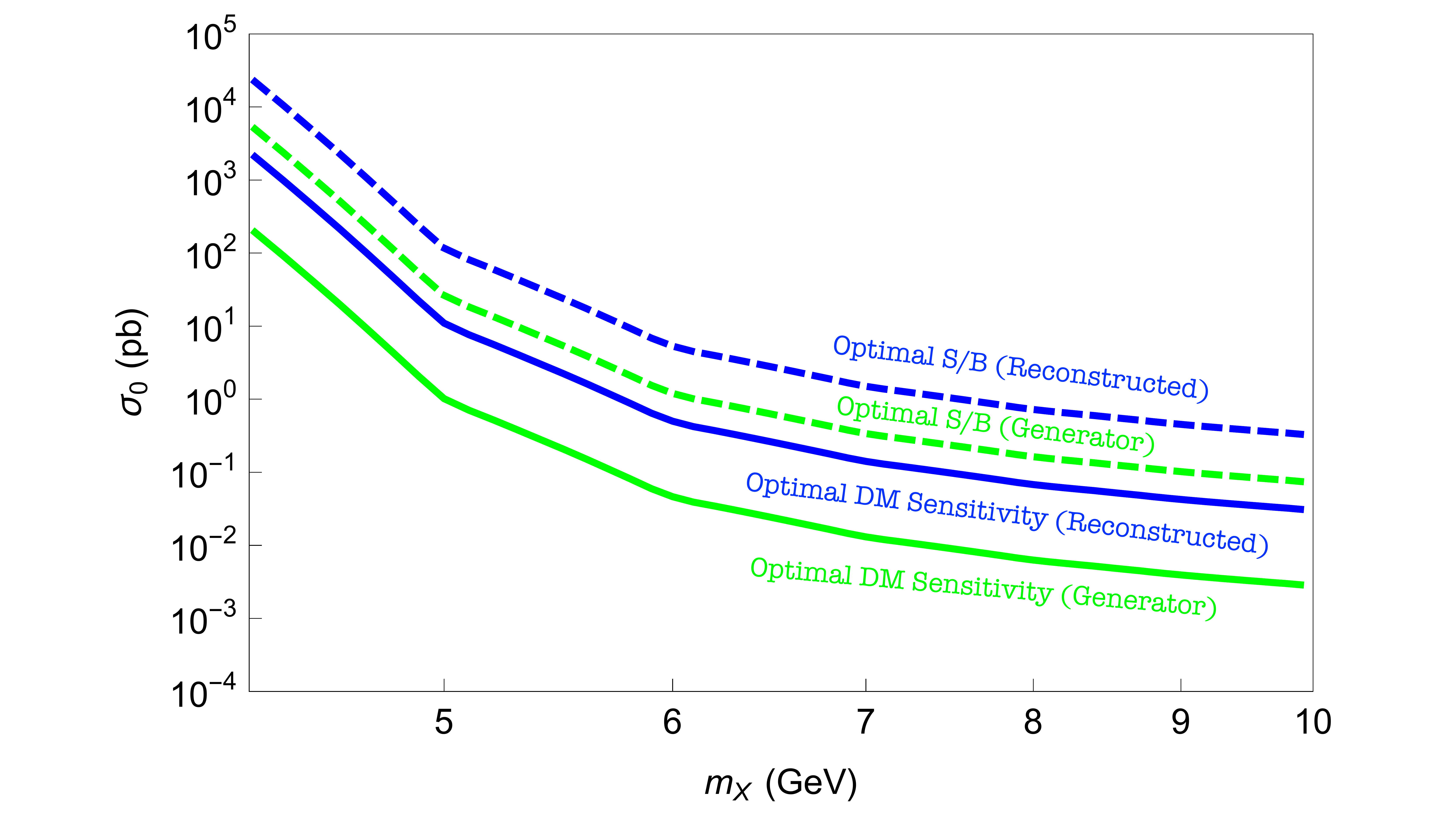}
  \caption{\label{fig:sensitivity} Projected 90\% sensitivity curves for DUNE (400 kT yr) for inelastic dark matter scattering. All the curves are for the stopped $K^{+}$ channel. The relevant cuts are listed in Fig.~\ref{fig:flux236} and Table~\ref{tab:Cuts}. }
\end{figure}

\textcolor{black}{In Figure~\ref{fig:sensitivity}, we} plot the 90\% CL sensitivity of DUNE (400 kT yr) in the $(m_{X}, \sigma_0)$-plane for the case where WIMPs annihilate solely to first generation quarks, assuming a search for monoenergetic neutrinos at 236 MeV from stopped $K^{+}$ decay.  
We plot sensitivity curves for each of the four cuts strategies 
given in Table~\ref{tab:Cuts}. 

%=============================================================================
%=============================================================================

\begin{table}[hbt!]
\caption{ The angular cuts (including the energy cut of $236 \pm 30 \mev$) and the resulting signal and background 
%fractions, 
\textcolor{black}{efficiencies,}
the expected number of signal and background events, the expected signal to background ratio at DUNE, and the 
maximum flux of 236 MeV neutrinos emanating from the Sun which would be allowed 
(at 90\% CL). The first two rows are cuts on reconstructed events and the last two rows are cuts on generator level events (no detector simulation/reconstruction). \textcolor{black}{We include the generator level information to illustrate the optimistic case of perfect reconstruction.}
\label{tab:Cuts}}
\vspace*{+6mm}
\centering
\begin{tabular}{|c|c|c|c|c|c|c|c|}
  \hline
  $\theta_{P}^{reco} < $ & $\theta_{N}^{reco} > $ & $\eta_S^{reco}$ & $\eta_B^{reco}$ & $N_{S}^{reco,90}$ & $N_{B}^{reco}$ & $S/B^{reco}$ & $\Phi_{236\mev}^{reco} [\m^{-2} \s^{-1}]$ \\
  \hline
&&&&&&&\\[-1em]
  $60^{\circ}$ & $162^\circ$ & $5.0 \times 10^{-4}$ & $1.9 \times 10^{-4}$ & 2.7 & 1.2 & 2.2 & $1.1 \times 10^{4}$ \\
  \hline
&&&&&&&\\[-1em]
    $60^{\circ}$ & $60^\circ$ & $2.7 \times 10^{-2}$ & $1.4 \times 10^{-2}$ & 13.9 & 92.7 & 0.2 & $1.1 \times 10^{3}$ \\
   \hline
  $\theta_{P} < $ & $\theta_{N} > $ & $\eta_S$ & $\eta_B$ & $N_{S}^{\ 90}$ & $N_{B}$ & $S/B$ & $\Phi_{236\mev} [\m^{-2} \s^{-1}]$ \\
  \hline
&&&&&&&\\[-1em]
    $50^{\circ}$ & $171^\circ$ & $1.8 \times 10^{-3}$ & $3.3 \times 10^{-5}$ & 2.1 & 0.2 & 10.5 & $2.4 \times 10^{3}$ \\
   \hline
&&&&&&&\\[-1em]
    $50^{\circ}$ & $20^\circ$ & $3.9 \times 10^{-1}$ & $2.6 \times 10^{-2}$ & 17.8 & 173.3 & 0.1 & 93.1 \\
   \hline
\end{tabular}
\end{table}

\textcolor{black}{
There are a variety of other theoretical uncertainties 
which can have a significant effect on DUNE's sensitivity.  
For example, we have assumed that dark matter annihilates 
to first generation quarks.  If dark matter annihilates 
instead to second generation quarks, the average number of 
$K^+$ produced per annihilation (and, thus, the flux of 
236 MeV neutrinos) would increase by about a factor of 2.  
Furthermore, we have modeled neutrino-nucleus scattering 
at this energy with} \verb+NuWro+.  
\textcolor{black}{Although there are 
experimental measurements of this cross section, there 
are still significant uncertainties, both in the 
magnitude of the charged-current cross section and in the 
angular dependence.  But any stopped pion experiment also acts 
as a stopped kaon experiment~\cite{Spitz:2012gp}, and a 
variety of future KDAR measurements 
are under consideration~\cite{Conrad:2010eu}, 
and would serve as a calibration 
for this type of analysis. Importantly, DUNE itself can 
provide calibration data, by searching off-axis.}

Future improvements in reconstruction techniques that could enable 
electron channel to be used effectively, would lead to a significant improvement in sensitivity.  
\textcolor{black}{The 
electron channel is generally expected to be more sensitive 
than the muon channel for three reasons~\cite{Rott:2016mzs}. 
First, the atmospheric 
neutrino background flux is smaller.  Second, the effective area of DUNE is larger for 236 MeV $\nu_e$ than for $\nu_\mu$,
because the charged-current scattering cross section for 
$\nu_\mu$ is suppressed by the reduced phase space of the 
outgoing muon.  Third, the flux of 236 MeV $\nu_e$ 
arriving at Earth from KDAR in the Sun is expected to 
be larger than the flux 
of 236 MeV $\nu_\mu$ as a result of oscillation effects in 
the dense medium of the Sun (assuming a normal hierarchy)~\cite{Lehnert:2007fv}.}

%\newpage
\section{Conclusion}

In this work, we have estimated DUNE's potential to detect 
the monoenergetic 236 MeV neutrinos arising 
from kaon-decay-at-rest in the core 
of the Sun. Although the 
charged leptons produced from a charged-current interaction 
of a 236 MeV neutrino are roughly isotropic, many such 
interactions produce an ejected proton which is 
forward-directed.  Moreover, the remnant nucleus 
tends to be backward-directed, and observable 
kinematic variables can be used as a proxy for the 
remnant nucleus momentum, allowing for better discrimination 
of signal from background.

We have used these observables in a realistic manner, with the response of the detector modelled 
numerically.  Although we have found that the discrimination of signal from background, $S/B$, can be as large as 2.2 for a model where there are enough signal events to exclude, a realistic treatment of the detector results in reduced sensitivity with respect to
earlier estimates.

Foreseeing future improvements in reconstruction (for example, via machine learning), we calculated the expected number of signal and background events which pass our cuts at the generator level (see Table~\ref{tab:Cuts}). We've also plotted the generator level dark matter sensitivity curves in green in Fig.~\ref{fig:flux236}. These are the limits in the optimistic case of perfect reconstruction, and we find 
that this optimal sensitivity matches estimates made 
previously~\cite{Rott:2016mzs}.

There are a variety of non-standard scenarios for dark matter 
particle physics and astrophysics in which the sensitivity of direct detection experiments is suppressed, 
and the flux of 236 MeV neutrinos produced in the Sun's 
core may provide an excellent indirect probe of dark matter 
interactions.  In this case, DUNE's ability to identify 
236 MeV neutrinos arriving from the direction of the 
Sun, while rejecting background, can provide unique 
control over systematic uncertainties.  As an example, we have 
estimated DUNE's sensitivity to low-mass dark matter which 
scatters inelastically, with a mass splitting of $\delta = 50 \kev$.  
This is an example of a dark matter process which is kinematically inacessible for 
direct detection experiments on Earth, but for which  a search for 
neutrinos at DUNE may lead to a discovery.  

The search for direct evidence of non-gravitational 
interactions between dark matter and Standard Model 
matter has thus far yielded no conclusive positive 
signals.  This has led to broader theoretical and experimental approaches to dark matter searches, 
and KDAR neutrinos can play an important 
role.  It would be interesting to further study the 
theoretical scenarios in which searches for KDAR 
neutrinos provide a competitive advantage.  

On the experimental side, it would 
also be interesting to study in more detail how the particle 
identification and track reconstruction at DUNE could 
be improved in the energy range relevant for KDAR 
searches. \textcolor{black}{A possible DUNE module-of-opportunity may use a wireless design with an isotropic response and could improve the sensitivity 
to dark matter annihilation in the Sun,  
by reducing the loss of efficiency associated with the 
orientation of the Sun with respect to the DUNE wires.}

%\newpage
\acknowledgments
% Standard piece for all FNAL-based experiments
This document was prepared by the DUNE collaboration using the
resources of the Fermi National Accelerator Laboratory 
(Fermilab), a U.S. Department of Energy, Office of Science, 
HEP User Facility. Fermilab is managed by Fermi Research Alliance, 
LLC (FRA), acting under Contract No. DE-AC02-07CH11359.
%
% Funding agencies, alphabetical by country, then alphabetical by agency name
%
This work was supported by
CNPq,
FAPERJ,
FAPEG and 
FAPESP,                         Brazil;
CFI, 
IPP and 
NSERC,                          Canada;
CERN;
M\v{S}MT,                       Czech Republic;
ERDF, 
H2020-EU and 
MSCA,                           European Union;
CNRS/IN2P3 and
CEA,                            France;
INFN,                           Italy;
FCT,                            Portugal;
NRF,                            South Korea;
CAM, 
Fundaci\'{o}n ``La Caixa'',
Junta de Andaluc\'ia-FEDER, and 
MICINN,                         Spain;
SERI and 
SNSF,                           Switzerland;
T\"UB\.ITAK,                    Turkey;
The Royal Society and 
UKRI/STFC,                      United Kingdom;
DOE and 
NSF,                            United States of America.
We are grateful to Xerxes Tata for useful discussions. C. Rott acknowledges support from the National Research Foundation of Korea.
%%%%%%%%%%%%%%%%%%%%%%%%%%%%%%%%%%%%%%%%%%%%%%%%%%%%%

%%%%%%%%%%%%%%%%%%%%%%%%%%%%%%%%%%%%%%%%%%%%%%%%%%%%%%
\end{document}